# How Big Can Anomalous W Couplings Be?

C.P. Burgess,[a]* M. Frank,[b] and C. Hamzaoui[c]

[a] *Institut de Physique, Université de Neuchâtel*
*1 Rue A.L. Breguet, CH-2000 Neuchâtel, Switzerland.*

[b] *Physics Department, Concordia University*
*1455 de Maisonneuve Blvd., Montréal, Québec, Canada, H3G 1M8.*

[c] *Département de Physique, Université du Québec à Montréal*
*Case Postale 8888, Succ. Centre-Ville, Montréal, Québec, Canada, H3C 3P8.*

**Abstract**

Order-of-magnitude estimates have given us the prevailing wisdom that anomalous electroweak boson self-couplings must be 1% or smaller for any reasonable new physics, essentially precluding their observation at LEP 200. We examine these estimates using the (for technical reasons, CP-violating) one-loop form factors that arise in a completely generic theory of spin-one, spin-half and spin-zero particles. We find form factors which can be in the range $(1-5)\%$ (for $q^2 \sim 4M_W^2$) and, due to threshhold enhancements, these largest values typically arise for light new physics, $m \lesssim 100$ GeV, where effective lagrangian analyses do not apply. For the case where the underlying physics does involve only very heavy particles these calculations are compared with the general effective-lagrangian description. We also identify the generic features of these moments, and find robust patterns through which the nature of the underlying physics becomes imprinted through the kinds of form factors that are generated.

* Permanent Address: *Physics Department, McGill University, 3600 University St., Montréal, Québec, Canada, H3A 2T8.*

## 1. Introduction

The last decades have established the Standard Model (SM) as the definitive description of all non-gravitational experiments. Although at any given time experiments could be found which would imply its imminent failure, all such challenges have (so far) disappeared on closer inspection. In this respect the SM has been like Mark Twain: news of its demise has proven to be greatly exaggerated.

Even so, the prevailing wisdom in the field is that the SM must soon fail. The big question is, where will it fail first? There are two types of experiments which have good discovery potential: those that directly explore higher energies, in the hopes of directly producing new particles; or those that examine previously unprobed types of interactions, where deviations from SM predictions may lurk.

The study of $e^+e^-$ collisions, at LEP 200, above the threshhold for $W^+W^-$ pair production provides an example of this second type. The self-couplings of the electroweak gauge bosons can be probed here, permitting the detection of nonstandard three-point interactions (TGV's) [1], [2], potentially down to the 10% level [2], [3]. Similar precision might also be achievable using the reaction $W \to W\gamma$ at the TeVatron, [4], and the LHC may ultimately present even better prospects [5], should it be built.

Of course, in order to be detected, an anomalous TGV at this level must be possible for some underlying physics, and it must have escaped detection elsewhere in other precision electroweak experiments. And although this is possible at the purely phenomenological level [6], [7], [8], there are reasonably persuasive theoretical arguments that indicate that TGV's are unlikely to be larger than 1% or less. These arguments proceed along several lines.

The simplest such argument simply amounts to coupling-constant counting. Loops involving new physics generate anomalous TGV's with a size of order $(g/4\pi)^2$, with $g$ an electroweak gauge coupling, and so are at most a few times $10^{-3}$. Since the gauge coupling is universal, this estimate includes a large class of models. One might hope to evade this bound by using the fact that longitudinal gauge bosons can couple with a strength that is larger than $g$. For example, in multihiggs models the relevant coupling is $gm_{H}/M_{W}$, which can be large if the Higgs is heavy, $m_{H} \sim 1$ TeV. Even in this strongly-coupled case reasonably persuasive arguments limit the potential size of the anomalous TGV's that can be generated [9]. Using power counting techniques that are motivated by experience with chiral perturbation theory in QCD, TGV's can be expected to be of order $M_{W}^2/m_{H}^2$ or smaller — *i.e.* at most a few percent — even if induced by TeV-scale scalars.

A separate argument to the same end is based on naturalness [6], [9]. Precision electroweak measurements already bound deviations from the SM in the $W^{\pm}$- and $Z^0$-boson vacuum polarizations (the 'oblique corrections') at the percent level [10], [11]. According to this line of reasoning, in the absence of a symmetry which distinguishes the oblique corrections from the TGV's, any new physics which generates one must also generate the other. If true, the better precision that is available for detecting oblique corrections would imply that any new physics must first be discovered there.

In the present paper we consider TGV's that are induced at one loop within a completely general renormalizable theory of spinless, spin-half and spin-one particles. We do so with the following motives.

• *Motive 1:* Our first goal is to use the explicit calculation as a vehicle for exploring the above theoretical arguments more systematically, and in more detail. Although the TGV's we compute are typically a few % in size, it is nevertheless still useful to know whether and under what circumstances these arguments fail. After all, if 10% effects are now expected to be detectable, perhaps 5% deviations or less will ultimately be visible once the experimenters have familiarized themselves with the new environment, given that they have sufficient motivation to look. If so, it would be of great interest to know more accurately just how large TGV's can be, and precisely which properties of the underlying theory are required to maximize their size. Indeed, in what follows we identify circumstances for which induced form factors can plausibly be in the range $(1-5)\%$.

Furthermore, potential loopholes in the above arguments do, in fact, exist. For example, arguments based on the properties of an effective lagrangian presuppose that the new physics is heavy enough to justify such an approach. It is conceivable, however, that the new physics is *not* much heavier than several hundred GeV, and yet has still escaped detection. We can address this possibility since we do not assume that our underlying new particles are heavy, and we compute the entire TGV *form factors* rather than simply their values at zero momentum transfer. We find that the largest effects do arise for light particles, where threshhold effects can enhance the sizes of the relevant one-loop results.

Similarly, the naturalness argument which states that new physics must appear first through oblique corrections before it can appear in TGV's, need not apply when the dominant couplings of the new physics violate $CP$. Since the $W^{\pm}$ and $Z^0$ vacuum polarizations cannot violate $CP$, $CP$ itself is a symmetry which can suppress the contributions of new physics to oblique corrections relative to its contribution to TGV's. For example, if $CP$-violating phenomena are suppressed by some small(ish) parameter, $\epsilon$, in an underlying theory, then $CP$-violating TGV's arise at $O(\epsilon)$, while their contributions to oblique corrections are $O(\epsilon^2)$. (Of course, the existence of this particular loophole may be cold comfort, since suppression by a small parameter like $\epsilon$ would make the induced TGV's even smaller than the expected O(1%). And any $CP$-preserving new-physics couplings would still need to escape detection.)

• *Motive 2:* Our second motivation takes a separate tack. Eventually TGV's will be observed, if not at LEP then perhaps elsewhere. Once they are seen (if they are anomalous and are unaccounted for by the particles that are then known), the paramount question will concern the diagnosis of the kind of new physics that might be responsible. In order to provide a diagnosis we need to know the systematics of how different types of underlying physics differ in the TGV's that they can induce. By working with a very general underlying model, we are able to determine some of these systematics, which we list in Section (5). In passing, we also correct some errors in the literature concerning the implications of custodial symmetries for these vertices.

For simplicity we restrict our attention here to $CP$-violating TGV's. Our main reason for doing so is that this eases the evaluation of the relevant Feynman graphs, although the extension to the general case is straightforward. This is also the case that is less well studied in the literature. We expect our results for the sizes of the induced form factors to apply equally well for the $CP$-invariant case.

Here is how we organize our presentation. After briefly reviewing the potential $CP$-violating three-point gauge boson self interactions in Section (2), we outline in Section (3) what might be expected on general grounds should the new physics which generates them be massive enough to have its effects be well described by an effective low-energy lagrangian. In Section (4) we identify the general features that follow for the electroweak form factors at one loop, and show how various classes of underlying physics can robustly leave their imprint through the pattern of TGV's which they produce. In Section (5) we then compute the $CP$-violating three-point interactions using more specific models which illustrate the potential sizes that can be obtained by supplementing the SM by additional spin-zero and spin-half particles. Using the underlying parameters which maximize the induced TGV's, in Section (6) we find representative underlying models which do not conflict with any other extant bounds. Our main conclusions are finally summarized in the final section.

## 2. $W^{\pm}$ and $Z^0$ Electroweak Moments

We first review the definitions of the various electroweak form factors, in order to establish our notation. There are three types of form factors that describe the $CP$-violating couplings of an (on shell) $W^{\pm}$ boson to the photon and three more describe the analogous $W^{\pm}$ couplings to the $Z^0$. They are given in terms of the matrix elements [1], [2] of either the electromagnetic current, $J^\mu_{\rm em}$, or the weak neutral current, $J^\mu_{\rm nc}$, by:

$$\langle W^-|J^\mu_{_V}|W^-\rangle \equiv -ie_{_V}\ \varepsilon^*_\alpha(p_1)\Gamma^{\alpha\beta\mu}_{_V}(p_1,p_2)\varepsilon_\beta(p_2), \tag{1}$$

where:

$$\Gamma^{\alpha\beta\mu}_{_V}\Big|_{\rm CP-odd} = \Gamma^{\alpha\beta\mu}_{_V}\Big|_{\rm P,CP-odd} + \Gamma^{\alpha\beta\mu}_{_V}\Big|_{\rm C,CP-odd},$$

and:

$$\begin{aligned}
\Gamma^{\alpha\beta\mu}_{_V}\Big|_{\rm P,CP-odd} &= -f_{_V}(q^2)\epsilon^{\alpha\beta\mu\rho}q_\rho - \frac{1}{M^2_{_W}}g_{_V}(q^2)\ p^\mu\epsilon^{\alpha\beta\sigma\rho}q_\sigma p_\rho \\
\Gamma^{\alpha\beta\mu}_{\gamma}\Big|_{\rm C,CP-odd} &= \frac{i}{M^2_{_W}}h_\gamma(q^2)\left[q^2\left(q^\alpha\eta^{\beta\mu}+q^\beta\eta^{\alpha\mu}\right)-2q^\mu q^\alpha q^\beta\right], \\
\Gamma^{\alpha\beta\mu}_{_Z}\Big|_{\rm C,CP-odd} &= ih_{_Z}(q^2)\left(q^\alpha\eta^{\beta\mu}+q^\beta\eta^{\alpha\mu}\right).
\end{aligned} \tag{2}$$

Only the $CP$-violating part of the matrix element is listed here explicitly. $p = p_1 + p_2$ and $q = p_1 - p_2$ are respectively the sum and difference of the four-momenta, $p_i$, of the initial and final $W$'s and $\varepsilon_\mu(p_i)$ are the $W$-boson polarization vectors. The subscript $V = \gamma$ or $Z$ indicates whether it is the electromagnetic or neutral-current coupling that is being considered. The coupling constant, $e_{_V}$, appropriate to each is taken as $e_\gamma = e$ and $e_{_Z} = e\,c_w/s_w$ in which $e$ is the proton's charge and $s_w$ ($c_w$) is the sine (cosine) of the electroweak mixing angle. Our notation here is related to that of Ref. [2] by: $f_{_V}(q^2) = f_6^V(q^2)$, $g_{_V}(q^2) = f_7^V(q^2)$, $(q^2/M^2_{_W})h_\gamma(q^2) = f_4^\gamma(q^2)$, and $h_{_Z}(q^2) = f_4^Z(q^2)$. We treat $h_{_Z}$ and $h_\gamma$ differently since the anomalous photon vertex is constrained by electromagnetic gauge invariance to be transverse, whereas the $Z$ couplings need not be.

For the $Z^0$ we have:

$$\begin{aligned}
\langle Z^0|J^\mu_{_C}|Z^0\rangle &\equiv -ie_{_Z}\ \varepsilon^*_\alpha(p_1)\hat\Gamma^{\alpha\beta\mu}_{_V}(p_1,p_2)\varepsilon_\beta(p_2), \\
\hat\Gamma^{\alpha\beta\mu}_{\gamma}\Big|_{\rm CP-odd} &= \frac{i}{M^2_{_Z}}\,\hat h_\gamma(q^2)\left[q^2\left(q^\alpha\eta^{\beta\mu}+q^\beta\eta^{\alpha\mu}\right)-2q^\mu q^\alpha q^\beta\right].
\end{aligned} \tag{3}$$

$$\hat\Gamma^{\alpha\beta\mu}_{_Z}\Big|_{\rm CP-odd} = \frac{i}{M^2_{_Z}}\,\hat h_{_Z}(q^2)\left(q^\alpha\eta^{\beta\mu}+q^\beta\eta^{\alpha\mu}\right). \tag{4}$$

The form factors $h_\gamma(q^2)$, $h_{_Z}(q^2)$, $\hat h_\gamma(q^2)$ and $\hat h_{_Z}(q^2)$ differ from the others in their transformation properties under $P$ and $C$. Unlike $f_{_V}(q^2)$ and $g_{_V}(q^2)$, which violate $P$ but preserve $C$, these effective interactions break $C$ but are $P$-even. These selection rules permit the form factors $f_\gamma(0)$ and $g_\gamma(0)$ to be much more strongly constrained than are the others. This is because they can contribute through loops to the electric dipole moments (*edm*'s) of elementary fermions, such as the electron and the neutron, for which there are extremely strong limits. The bound that is obtained in this way for $f_\gamma$ was originally derived in the second of Refs. [12], while an analysis for $g_\gamma$ and $h_\gamma$ may be found in Ref. [13]. The strongest bounds to emerge from these analyses are:

$$\begin{aligned} f_\gamma(0) &\lesssim O(10^{-3}), \\ g_\gamma(0) &\lesssim O(10^{-4}). \end{aligned} \tag{5}$$

Similar bounds do not hold for the other form-factors, which could be $O(1)$ so far as direct phenomenological bounds are concerned. The somewhat model-dependent exception to this statement arises if some of the remaining form factors are related to $f_\gamma$ or $g_\gamma$ by some symmetry of the low-energy theory, as can be true for $f_Z$ and $g_Z$, for instance [14]. It is perhaps worth emphasizing that $h_\gamma$ and $\hat{h}_\gamma$ are the only $CP$-violating electromagnetic vertices for which no bounds exist [15].

Due to the strength of the *edm* bounds it would be unnatural to expect to find any signal in any forseeable $e^+e^-$ collisions due to the electromagnetic form factors $f_\gamma$ and $g_\gamma$. (Recall that their detection at LEP or in hadron machines is only possible if they are not smaller than $\sim 0.1$ at the appropriate momentum transfers.) The same need *not* be true, however, for the other electroweak form factors. It is then irresistable to ask what kinds of new physics can produce them, how large they might reasonably be expected to be, and what general considerations can constrain their size. These are the topics of the following sections.

## 3. The Effective-Lagrangian Limit

Before computing these form factors in explicit models it is first worth extracting whatever information about them that may be had on general grounds. This section is devoted to determining the relations among the six electroweak form factors that follow simply from symmetry considerations if the new physics is associated with large energies, $M \sim 1$ TeV.[1]

In this case the characteristic scale over which the form factors vary appreciably is the large scale, $M$, and so they may be well approximated at current energies by their values at

[1] Part of our motivation for doing so here is to clear up a lingering confusion in the literature as to the implications of custodial symmetries for TGV's.

$q^2 \approx 0$. When this is true, the simplest way to infer the implications for these 'electroweak moments' is to write down the most general low-energy effective lagrangian [16] that is consistent with the assumed low-energy particle content and symmetries, and to compute the low-energy form factors within this theory. Any relations among the moments that follow from such a lagrangian then rely only on the assumed low-energy symmetries and spectrum, and must be shared by all underlying theories which possess these low-energy properties [17], [18].

Any such approach must make a fundamental choice: the particle content that is assumed to appear in the low-energy theory. In particular, a key issue is whether or not this low-energy particle content fills out a linear realization of the electroweak gauge group, $SU_L(2) \times U_Y(1)$. If so, then the low-energy lagrangian should include the fields whose vacuum expectation values ($v.e.v.$s) are responsible for electroweak breaking. These fields might consist of just the Standard-Model Higgs, or perhaps also other degrees of freedom as well.

If, on the other hand, it is the symmetry-breaking sector itself that has been integrated out then no Higgs field need appear in the low-energy lagrangian and the electroweak gauge group is nonlinearly realized. Equivalently, [19][20], only the electromagnetic part of the gauge group, $U_{\rm em}(1)$, need be linearly realized, and the rest of the electroweak gauge group may be ignored. An important feature which arises when the gauge group is nonlinearly realized, is that perturbative unitarity in the effective theory must break down at energies $E \gtrsim 8\pi M_W/g$. As a result, whatever new physics is responsible for the effective theory must involve mass scales that are smaller than this.

We consider, in turn, the implications of each of these choices in the following two subsections.

### *3.1) CP-Violation due to the Electroweak Breaking Sector*

Perhaps the least restrictive scenario has the new physics responsible for any measured $W$-boson moment be associated with the poorly-understood sector which breaks the electroweak gauge symmetries. In this case the effective lagrangian need only linearly realize $U_{\rm em}(1)$ invariance and so contains one term for each of the six possible form factors. As a result, no relations are required, *a priori*, among the form factors in this most general case.

The potential effective interactions are these. The effective interactions which have dimension (mass)$^4$ are conventionally written:

$$\mathcal{L}_{\rm four} = -ieW^*_\mu W_\nu \left( \tilde{\kappa}_\gamma \, \widetilde{F}^{\mu\nu} + \tilde{\kappa}_Z \, \frac{c_w}{s_w} \, \widetilde{Z}^{\mu\nu} \right) + e\frac{c_w}{s_w}(a_Z W^*_\mu W_\nu + \hat{a}_Z Z_\mu Z_\nu)(\partial^\mu Z^\nu + \partial^\nu Z^\mu), \tag{6}$$

where the 'tilde' represents the dual of the corresponding field strength: $\widetilde{Z}_{\mu\nu} = \frac{1}{2}\epsilon_{\mu\nu\lambda\rho} \, Z^{\lambda\rho}$. The rest have dimension six:

$$\mathcal{L}_{\rm six} = -ieW^{*\nu}_\mu \, W_\nu{}^\sigma \, \left( \tilde{\lambda}_\gamma \, \widetilde{F}_\sigma{}^\mu + \tilde{\lambda}_Z \, \frac{c_w}{s_w} \, \widetilde{Z}_\sigma{}^\mu \right) + e\,(a_\gamma W^*_\mu W_\nu + \hat{a}_\gamma Z_\mu Z_\nu)(\partial^\mu \partial_\lambda F^{\nu\lambda} + (\mu \leftrightarrow \nu)). \tag{7}$$

Notice that the terms proportional to $a_\gamma$ and $\hat{a}_\gamma$ can be rewritten in terms of the gauge-fermion and four-point gauge self-interactions using the SM equations of motion. This implies the complete equivalence of these two formulations of this interaction, since one can transform from one to another simply by performing a field redefinition [21]. We choose not to do so here, however, since we prefer to keep as much as possible of the universal contributions from new physics in the purely electroweak-boson sector.

The relation between the effective couplings in these effective interactions, and the form factors as defined earlier is:

$$\begin{aligned} f_\gamma(0) &= \tilde{\kappa}_\gamma + \tilde{\lambda}_\gamma \, M_W^2, \qquad & g_\gamma(0) &= -\frac{1}{2}\tilde{\lambda}_\gamma \, M_W^2, \qquad & h_\gamma(0) &= a_\gamma \, M_W^2 \\ f_Z(0) &= \tilde{\kappa}_Z + \tilde{\lambda}_Z \, M_W^2, \qquad & g_Z(0) &= -\frac{1}{2}\tilde{\lambda}_Z \, M_W^2, \qquad & h_Z(0) &= a_Z \end{aligned} \tag{8}$$

and

$$\hat{h}_\gamma(0) = \hat{a}_\gamma \, M_Z^2, \qquad \hat{h}_Z(0) = \hat{a}_Z. \tag{9}$$

To draw any further conclusions, we must make some further assumptions regarding the nature of the underlying physics which is responsible for these effective interactions. We now list a few of the most plausible assumptions, together with their implications for our anomalous moments. We do so in order to obtain some intuition as to what sizes are to be expected for different types of underlying physics. These expectations are compared with the results of our explicit calculations in later sections.

- *Power Counting:* The utility of any effective lagrangian relies on the validity of the low-energy expansion in powers of $E/M$, where $E$ is the energy of the phenomena which are to be described by the effective theory, and $M \gg E$ is the heavy mass which characterizes the scale of the underlying physics. This implies a heirarchy amongst the various effective interactions, although the precise nature of the heirarchy depends on the nature of the underlying theory [20].

If $E$ and $M$ were the only scales in the problem, then the effective theory could be expected to properly describe the $E/M$ expansion in the underlying theory if the coefficient of any effective interaction were $\lesssim O(1)$ times the power of $M$ which is required by dimensional analysis [22]. Applied to the TGV lagrangian of eq. (7), and including the a factor of the $SU_L(2)$ gauge coupling, $g$, for each $W$ boson, this gives the estimates in column one of Table I, under the label DA, for Dimensional Analysis.[2]

Unfortunately, things are not so simple for applications to $W$-boson scattering at several hundred GeV, as would be appropriate to LEP 200. In this case the electroweak symmetry-breaking scale, $v = 246$ GeV, arises, as well as the scales $E$ and $M$, and we are interested in the limit where $E \sim v \ll M$. As a result, if effective interactions are to properly reproduce the $1/M$ expansion in the underlying theory, then they may have to be additionally suppressed by powers of $v/M$ beyond the powers of $1/M$ that are required simply by dimensional analysis. (In practical examples this additional suppression factor typically turns out to be powers of $g^2 v^2/M^2 \sim M_W^2/M^2$.)

In particular, it is reasonable to expect our anomalous TGV couplings to become smaller as $M$ gets larger for fixed $v$ and $E$.[3] This property is automatic for the dimension-six interactions but, as may be seen from Table I, does not hold for the dimension-four operators unless their couplings are additionally suppressed by powers of $v/M$. An estimate of the size of this suppression can be made using the power-counting techniques — Naive Dimensional Analysis (NDA) [23] — of chiral perturbation theory that work well in QCD. These estimates are plausible since it may happen that the electroweak symmetry-breaking sector involves only TeV-scale states, and the longitudinal gauge bosons are strongly coupled. In this case, chiral perturbation theory would be expected to apply to the low-energy couplings of the would-be Goldstone bosons that make up these longitudinal modes. As applied to anomalous TGV's, the NDA estimate amounts once more to pure dimensional analysis, with the additional condition that both the dimension-four and dimension-six terms be suppressed by a factor of $v^2$. This estimate, together with the requirements of custodial symmetry (explained below) is displayed in column two of Table I.

It must be stressed that this is just an estimate, although it does have the advantage of systematically organizing all of the terms of the effective lagrangian into an expansion in powers of $v/M$ and $E/M$, applicable for $M \lesssim 4\pi v$. It need *not* include all possible underlying theories, nor need it necessarily provide an upper bound for the size that a coefficient in a generic underlying theory can take. This is illustrated in Table I, where column three gives an equally plausible estimate (explained below) which is based on different physical assumptions concerning the nature of the underlying physics. Although columns two and three are very similar to one another (if we identify $\varepsilon \sim v^2/M^2$), they disagree, for example, on their implications for the coefficients $\tilde{\lambda}_\gamma$ and $\tilde{\lambda}_Z$.

[2] The factors of $\varepsilon$ reflect the requirements of custodial symmetry, and are explained below.

[3] It must be kept in mind in making these arguments that $M$ cannot become larger than $O(4\pi v)$ without running into troubles with perturbative unitarity.

| Coupling | DA/Custodial | NDA/Custodial | LRDA |
|---|---|---|---|
| $\tilde{\kappa}_\gamma$ | $g^2$ | $g^2 v^2/M^2$ | $g^2 v^2/M^2$ |
| $\tilde{\kappa}_Z$ | $-\tilde{\kappa}_\gamma\, s_w^2/c_w^2 + \varepsilon\, g^2$ | $-\tilde{\kappa}_\gamma\, s_w^2/c_w^2 + \varepsilon\, g^2 v^2/M^2$ | $-\tilde{\kappa}_\gamma\, s_w^2/c_w^2 + g^2 v^4/M^4$ |
| $\tilde{\lambda}_\gamma$ | $g^2/M^2$ | $g^2 v^2/M^4$ | $g^2/M^2$ |
| $\tilde{\lambda}_Z$ | $g^2/M^2$ | $g^2 v^2/M^4$ | $\tilde{\lambda}_\gamma + g^2 v^2/M^4$ |
| $a_\gamma$ | $g^2/M^2$ | $g^2 v^2/M^4$ | $g^2 v^2/M^4$ |
| $a_Z$ | $\varepsilon\, g^2$ | $\varepsilon\, g^2 v^2/M^2$ | $g^2 v^4/M^4$ |
| $\hat{a}_\gamma$ | $g^2/M^2$ | $g^2 v^2/M^4$ | $g^2 v^2/M^4$ |
| $\hat{a}_Z$ | $\varepsilon\, g^2$ | $\varepsilon\, g^2 v^2/M^2$ | $g^2 v^4/M^4$ |

**Table I**

This table displays a comparison of the estimated sizes for each of the effective TGV couplings using the three scenarios, discussed in the text, for the underlying physics. DA stands for Dimensional Analysis, which consists of simply counting dimensions and powers of the gauge coupling, $g$. NDA and LRDA respectively denote better-motivated estimates, using 'Naive Dimensional Analysis' and 'Linearly-Realized Dimensional Analysis', as explained in the text.

• *Custodial Symmetry:* We do have some information beyond the simplest power counting, however. The relative strength of the low-energy neutral- and charged-current weak interactions, as described by the parameter $\rho$, agree with the tree-level SM result, $\rho = 1$, to within roughly a percent. This prediction is very sensitive to the nature of electroweak symmetry-breaking physics, and its natural explanation points to the existence of an approximate custodial $SU_c(2)$ symmetry [24]. Any such custodial symmetry has many more implications than simply the suppression of deviations from $\rho = 1$. It imposes a heirarchy on all effective interactions which involve the electroweak gauge bosons, in particular among the anomalous TGV's. Since the implications of custodial symmetry are often incorrectly stated in the literature, we derive them here in some detail.[4]

By its very definition, the custodial symmetry must enforce the vanishing of $\Delta\rho = \rho - 1$. This is acheived by requiring the three real fields, $W^a_\mu$ with $a = 1, 2, 3$, to transform as a $SU_c(2)$-triplet:

$$\delta W^a_\mu = \epsilon_{abc}\omega^b W^c_\mu, \tag{10}$$

where $\omega^b$ is an infinitesimal $SU_c(2)$ transformation parameter, $W_\mu = \frac{1}{\sqrt{2}}(W^1_\mu + iW^2_\mu)$ is the charged $W$-boson field, and $W^3_\mu = c_1 Z_\mu + c_2 A_\mu$ is an as-yet-unspecified linear combination of the propagating $Z$ and photon fields. The constants $c_1$ and $c_2$ are determined by diagonalizing the kinetic and mass terms, and will differ, in general, from the SM values. We take fermions to transform in a vectorlike way, with $\binom{u}{d}$ transforming as doublets, for each generation.

The implications of this custodial symmetry are obtained by writing out the most general lagrangian that it permits. We therefore demand a lagrangian of the following form:

$$\mathcal{L}(W, \psi, a) = \mathcal{L}_{\rm inv}(W, \psi) + \mathcal{L}_{\rm sb}(W, \psi) + \mathcal{L}_{\rm em}(W, \psi, a), \tag{11}$$

where $W$ denotes the three fields, $W^a_\mu$, $a_\mu$ is the electromagnetic gauge potential, and $\psi$ schematically represents all of the remaining matter fields. The three types of terms have the following explanation:

**1)** $\mathcal{L}_{\rm inv}$: These are the most general terms possible which do not involve the photon field, and which are custodial invariant. Since these terms preserve this symmetry, their coefficients can be $O(1)$ (as indeed they are in the Standard Model).

**2)** $\mathcal{L}_{\rm em}$: Since the electromagnetic interactions explicitly break the custodial symmetry – *e.g.* $u$ and $d$ have different electric charges, even in the Standard Model – we do not require that any terms involving the field $a_\mu$ be symmetric under $SU_c(2)$. Coupling constants here nevertheless need not be particularly suppressed beyond the powers of the electromagnetic coupling, $e$, which accompany factors of $a_\mu$.

**3)** $\mathcal{L}_{\rm sb}$: This contains all of the terms in the lagrangian which explicitly break the custodial symmetry but do *not* involve the electromagnetic gauge potential. Included are, in particular, contributions to $\Delta\rho$, so we demand that couplings in $\mathcal{L}_{\rm sb}$ must all be suppressed by small symmetry-breaking parameters, $\varepsilon \sim 0.01$. Such a suppression is completely natural because of the approximate $SU_c(2)$ symmetry. It is not ruined by radiative corrections arising from the interactions in $\mathcal{L}_{\rm em}$, since these contribute to $\mathcal{L}_{\rm sb}$ couplings of order $\alpha/4\pi$.

[4] Our treatment here follows a similar analysis that appeared in the unpublished version of Ref. [20].

The remainder of the argument now proceeds as follows. By diagonalizing the gauge kinetic and mass terms we first determine the constants $c_1$ and $c_2$ that give the overlap between $W^3$ and the propagation eigenstates, $Z$ and $A$. By considering the electroweak couplings of fermions, we next verify that the custodial symmetry as defined really does ensure $\rho = 1$ to lowest order. Finally, we compute the implications of the symmetry for anomalous TGV's.

The most general $U_{\rm em}(1)$-invariant free lagrangian which involves only the fields $W^a_\mu$ and $a_\mu$ may be written:

$$\begin{aligned}
\mathcal{L}_{\rm inv}^{\rm free} &= -\frac{1}{4} W^a_{\mu\nu} W_a^{\mu\nu} - \frac{1}{2} m_{\scriptscriptstyle W}^2 \; W^a_\mu W^\mu_a, \\
\mathcal{L}_{\rm sb}^{\rm free} &= -\frac{s}{4} \; W^3_{\mu\nu} W^{3\mu\nu} - \frac{t}{2} \, m_{\scriptscriptstyle W}^2 \; W^3_\mu W^{3\mu}. \\
\mathcal{L}_{\rm em}^{\rm free} &= -\frac{1}{4} f_{\mu\nu} f^{\mu\nu} - \frac{\lambda}{2} \, W^3_{\mu\nu} f^{\mu\nu},
\end{aligned} \tag{12}$$

where $f_{\mu\nu}$ and $W^a_{\mu\nu}$ are the usual abelian curls for the fields $a_\mu$ and $W^a_\mu$. (We choose to relegate the electromagnetic interactions which arise from the use of the $U_{\rm em}(1)$ gauge covariant derivative, $D_\mu W_\nu$, into $\mathcal{L}_{\rm em}$.) We have used the freedom to rescale the fields in an $SU_c(2)$-invariant way to put two of the kinetic terms into their standard forms. The approximate custodial symmetry leads us to expect $\lambda \sim O(1)$, while $s, t \sim O(\varepsilon)$.

The neutral-boson propagation eigenstates for this free lagrangian are:

$$Z_\mu \equiv \sqrt{1 + s - \lambda^2} \; W^3_\mu, \qquad A_\mu \equiv a_\mu + \lambda W^3_\mu. \tag{13}$$

and the vector-boson masses become:

$$-m_{\scriptscriptstyle W}^2 \; W^*_\mu W^\mu - \frac{1}{2} \left( \frac{1+t}{1+s-\lambda^2} \right) m_{\scriptscriptstyle W}^2 \; Z_\mu Z^\mu. \tag{14}$$

To check the size of $\Delta\rho$ we write down the boson-fermion couplings that are unsuppressed by $\varepsilon$:

$$\mathcal{L}_{\rm inv}^{\scriptscriptstyle GF} = ig \, \overline{\psi} \gamma^\mu T_a \gamma_{\scriptscriptstyle L} \psi \; W^a_\mu \qquad \text{and} \qquad \mathcal{L}_{\rm em}^{\scriptscriptstyle GF} = ie \, \overline{\psi} \gamma^\mu Q \psi \; a_\mu. \tag{15}$$

$T_a$ and $Q$ here respectively represent the generators of $SU_c(2)$ and $U_{\rm em}(1)$ on the fermion fields.[5]

Transforming to boson propagation eigenstates, we obtain the standard charged-current and electromagnetic interactions, with respective strengths given in the usual way by $g$ and $e$, together with the following neutral-current interaction:

$$\mathcal{L}_{\rm nc} = \frac{ig}{\sqrt{1+s-\lambda^2}} \; \overline{\psi}\gamma^\mu \left( T_3 \gamma_L - \frac{e\lambda}{g} Q \right) \psi \; Z_\mu . \tag{16}$$

From this we infer the SM values, $s_{SM} = t_{SM} = 0$, $g_{SM} = e/s_w$ and $\lambda_{SM} = s_w$, as well as the tree-level prediction for $\Delta\rho$:

$$\Delta\rho = -\frac{t}{1+t} \approx -t \sim O(\varepsilon), \tag{17}$$

as required. Although the variables $s$ and $\lambda$ drop out of $\Delta\rho$, they do contribute to precision electroweak measurements through their contributions to the oblique-correction parameters $S$, $T$ and $U$ [25]:

$$\frac{\alpha S}{4 s_w^2 c_w^2} = -s + \frac{c_w^2 - s_w^2}{s_w c_w} (\lambda - \lambda_{SM}), \quad \alpha T = -t, \quad \frac{\alpha U}{4 s_w^2 c_w^2} = s + 2\frac{s_w}{c_w} (\lambda - \lambda_{SM}). \tag{18}$$

The implications for $CP$-violating anomalous TGV's are equally simple to work out. TGV's that arise from $\mathcal{L}_{\rm inv}$ or $\mathcal{L}_{\rm em}$ need not be suppressed by small custodial-breaking parameters. The most general such terms up to dimension six are

$$\begin{aligned} \mathcal{L}_{\rm inv}^{TGV} &= \frac{g\,\tilde{\xi}}{3!}\, \epsilon_{abc}\, W_\mu^{a\nu}\, W_\nu^{b\sigma}\, \widetilde{W}_\sigma^{c\mu}, \\ \mathcal{L}_{\rm em}^{TGV} &= -ie\tilde{\kappa}_\gamma\, W_\mu^* W_\nu\; \tilde{f}^{\mu\nu} - ie\tilde{\lambda}_\gamma\, W_\mu^{*\nu}\, W_\nu{}^\sigma\; \tilde{f}_\sigma{}^\mu \\ &\qquad + e(a_\gamma\, W_\mu^* W_\nu + \hat{a}_\gamma Z_\mu Z_\nu)(\partial^\mu \partial_\lambda f^{\nu\lambda} + (\mu \leftrightarrow \nu)). \end{aligned} \tag{19}$$

Notice that the other potentially invariant combination, $\epsilon_{abc}\; W_\mu^a W_\nu^b\; \widetilde{W}^{c\mu\nu}$, turns out to be identically zero.

[5] We assume here a universal strength for the $W^a$ couplings to fermions, as well as no right-handed charged currents, neither of which follow automatically from custodial invariance. They can be ensured by enforcing other approximate flavour symmetries, but we ignore this complication as peripheral to our main point.

In terms of propagation eigenstates these expressions become:

$$\mathcal{L}^{TGV}_{\rm inv} = -\frac{ie\tilde{\xi}}{s_w c_w} W^{*\nu}_{\mu} W_{\nu}{}^{\sigma}\, \tilde{Z}_{\sigma}{}^{\mu} \tag{20}$$

$$\mathcal{L}^{TGV}_{\rm em} = -ie\tilde{\kappa}_\gamma\, W^*_\mu W_\nu\, \left(\tilde{F}^{\mu\nu} - \frac{s_w}{c_w}\, \tilde{Z}^{\mu\nu}\right) - ie\tilde{\lambda}_\gamma\, W^{*\nu}_{\mu} W_{\nu}{}^{\sigma}\, \left(\tilde{F}_{\sigma}{}^{\mu} - \frac{s_w}{c_w}\, \tilde{Z}_{\sigma}{}^{\mu}\right)$$
$$+ e(a_\gamma\, W^*_\mu W_\nu + \hat{a}_\gamma Z_\mu Z_\nu)(\partial^\mu \partial_\lambda F^{\nu\lambda} + (\mu \leftrightarrow \nu)).$$

where in these expressions we drop any small deviations in $s, t, g$ and $\lambda$ from their SM values, as well as a $q^2$-dependent contribution to $h_Z$ and $\hat{h}_Z$.

The rest of the anomalous $CP$-violating TGV's must be suppressed by at least one factor of the small symmetry-breaking parameter, $\varepsilon$. Notice that, although the custodial symmetry relates $\tilde{\kappa}_\gamma$ and $\tilde{\kappa}_Z$ according to $\tilde{\kappa}_Z = -\tilde{\kappa}_\gamma\, s^2_w / c^2_w$, no relation is imposed between $\tilde{\lambda}_\gamma$ and $\tilde{\lambda}_Z$.

The resulting factors of $\varepsilon$ that are predicted in this way for each of the anomalous TGV's are listed in Columns one and two of Table I.

### *3.2) $CP$-Violation NOT associated with the Electroweak Breaking Sector*

In the alternative scenario, the underlying physics which generates anomalous TGV's is not also responsible for electroweak symmetry breaking. In this case it is likely that the low-energy effective theory linearly realizes the electroweak gauge group, and so contains the fields whose *v.e.v.*s give masses to the $W$ and the $Z$. We suppose for the present purposes that these fields are scalars, and that these scalars transform as $SU_L(2)$ doublets, since in this case the $\rho$ parameter is naturally close to one. The resulting effective operators must then be constructed from these fields, and must be $SU_L(2) \times U_Y(1)$ invariant [26]. This is the case analysed in Refs. [14], [18] and we repeat the conclusions here for comparison with the previous section.

In this framework the lowest-dimension operators which contribute to $CP$-violating anomalous TGV's arise at dimension six. They may be written as a linear combination of the following two independent ones:

$$A\, \frac{g^3}{3!} \epsilon_{abc} W^{a\nu}_{\mu} W^{b\lambda}_{\nu} \widetilde{W}^{c\mu}_{\lambda} + gg' \,{\rm Re}\, \left[B_{ij}\, (\phi^\dagger_i t_a \phi_j)\right] W^a_{\mu\nu} \widetilde{B}^{\mu\nu}. \tag{21}$$

Here $t_a$ are the $SU_L(2)$ generators as represented on the Higgs fields, $g'$ and $g$ denote the gauge couplings for each of the two factors of the electroweak gauge group, whose field strengths are written $B_{\mu\nu}$ and $W^a_{\mu\nu}$. $\phi_i, i = 1, \dots, N$ are the Higgs doublets, whose (potentially complex) $v.e.v.$s we denote $\langle \phi_i \rangle = \binom{0}{w_i}$. (In the case of a single Higgs, $w = v/\sqrt{2}$.)

Unlike the previous case, power counting arguments for this effective theory merely require that the coefficients $A$ and $B_{ij}$ be $O(1/M^2)$, since all additional suppression by powers of $w_i$ (or $v$) arises from the explicit dependence on the fields $\phi_i$.

Once evaluated with $\phi_i = \langle \phi_i \rangle$, these effective interactions produce the following effective TGV couplings:

$$\tilde{\kappa}_\gamma = -g^2 \, \mathrm{Re} \, \left( B_{ij} w_i^* w_j \right), \qquad \tilde{\kappa}_Z = g^2 \, \mathrm{Re} \, \left( B_{ij} w_i^* w_j \right) \frac{s_w^2}{c_w^2},$$
$$\tilde{\lambda}_\gamma = g^2 \, A, \qquad \tilde{\lambda}_Z = g^2 \, A. \tag{22}$$

Notice that $\tilde{\kappa}_\gamma$ and $\tilde{\kappa}_Z$ are related to one another in the same way as is required by the custodial symmetry, while the relation $\tilde{\lambda}_\gamma = \tilde{\lambda}_Z$ is *not* attributable to $SU_c(2)$ invariance. Being generated by dimension-six operators, the corresponding form factors, $g_V(0)$, are expected to be $O(M_W^2/M^2)$ in size. Deviations from the above relations amongst the form factors arise once higher-dimension operators are included, and so should be $O(M_W^4/M^4)$.

These estimates for the sizes of the $CP$-violating form factors are summarized in column three – entitled 'LRDA' for 'Linearly-Realized Dimensional Analysis' – of Table I. As was emphasized earlier, they do *not* necessarily agree with the estimates obtained from NDA.

### 4. Generic Features at One Loop

We next turn to the explicit one-loop calculation of the six form factors. For the purposes of the present section, we take as our underlying theory a completely general renormalizable theory of interacting spinless, spin-half and spin-one particles. In this way we can identify those features of the induced anomalous TGV's which are generic to differing kinds of underlying physics. The main signatures we find in this way are summarized in Table II, below.

We accordingly start with the most general renormalizable lagrangian that is possible for this particle content. Without loss of generality we choose to represent the fermions using a basis of Majorana spinors, $\psi^i$, just as we can (and do) represent our spinless particles with real scalar fields, $\phi_a$. The spin-one gauge potentials are $A^\alpha_\mu$. The lagrangian

is:

$$\mathcal{L} = -\frac{1}{2}\,(D_\mu\phi)_a(D^\mu\phi)_a - \frac{1}{2}\,\overline{\psi}^i \not{D}\psi^i - \frac{1}{4}\,F^\alpha_{\mu\nu}\,F^{\alpha\mu\nu} - V(\phi)$$
$$-\frac{1}{2}m_{ij}\,\overline{\psi}^i\gamma_L\psi^j + \frac{1}{2}\Gamma^a_{ij}\,\overline{\psi}^i\gamma_L\psi^j\;\phi_a + \text{h.c.} \tag{23}$$

Here $D_\mu\phi = \partial_\mu\phi - iA^\alpha_\mu\,t_\alpha\phi$ and $D_\mu\psi = \partial_\mu\psi - iA^\alpha_\mu\,(T_a\gamma_L - T^*_a\gamma_R)\psi$ are the gauge covariant derivatives, and $F^\alpha_{\mu\nu} = \partial_\mu A^\alpha_\nu - \partial_\nu A^\alpha_\mu + c^\alpha{}_{\beta\gamma}A^\beta_\mu A^\gamma_\nu$ is the covariant field strength. The matrix representation of the hermitian generators of the gauge group acting on the scalar fields is denoted $t_\alpha$, and satisfy the algebra $[t_\alpha, t_\beta] = ic^\gamma{}_{\alpha\beta}\,t_\gamma$. For real scalar fields we have $t_\alpha = t^\dagger_\alpha = -t^*_\alpha$. The matrix representation of the gauge generators acting on left-handed fermions is similarly denoted by $T_\alpha$, and satisfy the identical algebra. We adopt a normalization in which the gauge couplings, $g$, are absorbed into the definitions of our gauge generators, $t_\alpha$ and $T_\alpha$. The scalar potential can be a generic quartic polynomial: $V = \frac{1}{2}(\mu^2)^{ab}\phi_a\phi_b + \frac{1}{3!}\xi^{abc}\phi_a\phi_b\phi_c + \frac{1}{4!}\lambda^{abcd}\phi_a\phi_b\phi_c\phi_d$, subject to the gauge symmetries. Finally, the matrices $m_{ij}$ and $\Gamma^a_{ij}$ are in general complex, must preserve the gauge symmetries, and can be taken to be symmetric in their fermion indices $i$ and $j$.

Once the scalar potential has been minimized, the scalar fields, $\phi_a$, acquire their *v.e.v.*s $\langle\phi_a\rangle = v_a$. Perturbing around these values using $\phi_a = v_a + \chi_a$, and working within unitary gauge, gives the scalar, spinor and gauge-boson mass matrices. These are, respectively: $(M^2_S)^{ab} = (\mu^2)^{ab} + \xi^{abc}v_c + \frac{1}{2}\lambda^{abcd}v_cv_d$, $(M_F)_{ij} = m_{ij} + \Gamma^a_{ij}v_a$, and $(M^2_V)_{\alpha\beta} = \frac{1}{2}v^T\{t_\alpha, t_\beta\}v$. We choose to formulate our Feynman rules within a basis of propagation eigenstates, and so we redefine our fields to diagonalize these mass matrices. This has the effect of altering all of the couplings of the theory $t_\alpha \to \hat{t}_\alpha$, *etc.*, although we do not write the 'caret's in what follows, for brevity of notation.

We now turn to the calculation of one-loop TGV's from these interactions. Since our theory is renormalizable, and no divergent subdiagrams are possible at one loop, we are guaranteed that all anomalous form factors must be finite. As a result we need not be concerned here with the intricacies of various regularization schemes.

*4.1) Gauge Bosons*

The simplest interactions to deal with are those which couple the gauge bosons to one another:

$$\mathcal{L}_{GB} = -c_{\alpha\beta\gamma}\;\partial_\mu A^\alpha_\nu A^{\beta\mu}A^{\gamma\nu} - \frac{1}{4}c_{\alpha\beta\gamma}c^{\alpha\delta\epsilon}\;A^\beta_\mu A^\gamma_\nu A^\mu_\delta A^\nu_\epsilon. \tag{24}$$

These can, in principle, contribute to anomalous TGV's through the diagrams of Fig. (1). These diagrams turn out to never generate $CP$-violating interactions, however, as may be seen by the following simple argument.

The main point is that it is always possible to assign $C$ and $P$ quantum numbers to the gauge bosons in such a way as to ensure that the interactions of eq. (24) by themselves preserve $C$ and $P$. Since we use real fields here, one such assignment is to simply choose all fields, $A^\alpha_\mu$ to be eigenstates of $C$ and $P$ with eigenvalues $C = P = +1$.

Of course, once other interactions are included these $C$ and $P$ assignments may not turn out to preserve $CP$ in these other interactions. But this is irrelevant for the purposes of generating a $CP$-violating TGV from Fig. (1), since this graph knows only about the gauge self-couplings. It is therefore sufficient, for the purposes of eliminating $CP$-violating TGV's, to establish that there exists a set of $C$ and $P$ charge assignments for which the gauge self-couplings alone can be made to preserve $C$ and $P$.

*4.2) Fermions*

The only fermion interaction that can contribute to anomalous TGV's at one loop are the fermion-gauge couplings:

$$\mathcal{L}_{FG} = \frac{i}{2}\; \overline{\psi} \gamma^\mu (T_\alpha \gamma_L - T^*_\alpha \gamma_R) \psi \; A^\alpha_\mu. \tag{25}$$

These interactions can contribute to TGV's *via* the Feynman graph of Fig. (2). Special cases of this graph have been evaluated for left-right-symmetric models in Refs. [14], [27] and [28]. Since we work with propagation eigenstates, $CP$ violation enters into the graph through the vertices. These violate $CP$ if their coupling matrices, $T_\alpha$, are complex – *i.e.* not purely real or imaginary. Since these matrices are hermitian, $CP$-violation therefore can only arise within the off-diagonal terms in eq. (25).

It might be expected that since both $f_V(q^2)$ and $g_V(q^2)$ share the same transformation properties under $C$ and $P$, both would be generated whenever either is. Interestingly enough, this turns out not to be the case. Explicit calculation of Fig. (2) shows that both $g_\gamma(q^2)$ and $g_Z(q^2)$ *vanish* identically at one loop! The one-loop integrand is just not complicated enough to generate the required tensor structure, so these form factors never arise. This is an artifact of the one-loop approximation, however, and is no longer true at two loops [29].

More specific information may be extracted by considering Fig. (2) in more detail. If the neutral-boson vertex is diagonal in flavour space, as is necessarily true for the photon, then, as we have argued, it cannot break $CP$. In this case all of the $CP$ violation must come from phases in the fermion-W couplings. But in order for these phases not to cancel between the two W vertices in Fig. (2), there must be two independent couplings between the participating fermions and the $W$ boson. Since the photon couplings cannot change particle type, the only nonzero possibility arises when the fermions in the loop have both left- and right-handed couplings to the $W$ boson. In this case a $CP$-violating result must be proportional to the product of the left- and right-handed couplings, and so (since the photon couplings preserve $C$ and $P$) the result necessarily also breaks $P$. The conclusion, then – as was pointed out in Ref. [14] – is that the only $CP$-violating $WW\gamma$ form factor that can be produced by Fig. (2) is $f_\gamma(q^2)$, and this can arise only if the fermions have both left- and right-handed couplings to the $W$ boson.

This same arguments do not so strongly restrict the neutral-current moments, such as $WWZ$ or $ZZZ$. In this case more possibilities exist because the $Z$-fermion vertex can change flavour and can itself be $P$- and/or $CP$-violating. As a result, for instance, $f_Z(q^2)$ can be nonzero even in the absence of right-handed charged currents, provided there are instead flavour-changing $Z$-fermion interactions. Similarly, $h_Z(q^2)$ and/or $\hat{h}_Z(q^2)$ can also be nonzero, although the same considerations in this case imply that these form factors must be antisymmetric (at one loop) under the interchange of the masses of the two fermions which couple to the $Z$.[6] (On this point our results appear to differ with those of Ref. [28].)

These conclusions are summarized in column two of Table II.

*4.3) Scalars*

There are several gauge-scalar interactions that can contribute to one-loop TGV's. They are:

$$\mathcal{L}_{GS} = \mathcal{L}_{\chi\chi A} + \mathcal{L}_{\chi AA} + \mathcal{L}_{\chi\chi AA} \tag{26}$$

where

$$\begin{aligned} \mathcal{L}_{\chi\chi A} &= -\frac{i}{2} \; (\chi^T t_\alpha \partial^\mu \chi - \partial^\mu \chi^T t_\alpha \chi) \; A^\alpha_\mu \\ \mathcal{L}_{\chi AA} &= -\frac{1}{2} \; (v^T \{t_\alpha, t_\beta\} \chi) \; A^\alpha_\mu A^{\beta\mu} \\ \mathcal{L}_{\chi\chi AA} &= -\frac{1}{4} \; (\chi^T \{t_\alpha, t_\beta\} \chi) \; A^\alpha_\mu A^{\beta\mu}. \end{aligned} \tag{27}$$

[6] That is, they must be antisymmetric with respect to $m_i \leftrightarrow m_j$ in the notation of Fig. (2).

| Form Factor | Gauge Bosons | Fermions | Scalars |
|---|---|---|---|
| $f_V(q^2)$ | 0 | Nonzero | 0 |
| $g_V(q^2)$ | 0 | 0 | 0 |
| $h_V(q^2)$ | 0 | $Z$ Only | $Z$ Only |
| $\hat{h}_V(q^2)$ | 0 | $Z$ Only | $Z$ Only |

**Table II**

This table displays a comparison of the estimated sizes for each of the effective $CP$-violating TGV form factors that can arise at one loop. Each column represents the general result of the effects of couplings of the corresponding type of particles to gauge bosons. Items labelled '$Z$ Only' are skew under the interchange $m_i \leftrightarrow m_j$ (see text), and so necessarily vanish when $V{=}\gamma$.

In this case there are a great many types of graphs which can, in principle, contribute to anomalous TGV's. These are listed in Figs. (3) through (6). Some of these have been considered within the context of a two-Higgs-doublet model in Refs. [28] and [20].

It is possible to draw some general conclusions concerning these interactions, along the lines of those which were obtained for fermion loops in the previous section. In particular, it is possible to choose parity assignments for each of these interactions in such a way as to make them all $P$ invariant. This may be done by choosing all fields to be parity even (that is to say: scalars and polar vectors rather than pseudoscalars and axial vectors). As a result it is impossible to generate either of the form factors $f_V(q^2)$ or $g_V(q^2)$ from any of these scalar graphs. The only possible $CP$-violating contributions are therefore to $h_V(q^2)$ and to $\hat{h}_V(q^2)$. This gives $CP$-violation in the scalar sector a distinctive signature.

We see that $CP$-violating TGV form factors have a natural diagnostic property. Which form factors arise directly reflects the nature of the underlying physics which is responsible. In particular, $CP$-violation among new heavy scalars can only generate anomalous interactions which break both $CP$ and $C$, while underlying fermions can instead produce both $C$- and $P$-breaking $CP$-violating couplings. Neither type can generate a nonzero $g_V(q^2)$ at one loop, however.

As before, more detailed information is also available. For electromagnetic vertices, $CP$-violation must enter the graph through the scalar couplings to the $W$ boson. Since the electromagnetic couplings cannot change scalar flavour, and since only one type of scalar-$W$ vertex is possible for each flavour, $CP$-violation cancels between the two $W$ vertices in the photon moments. $h_\gamma(q^2)$ and $\hat{h}_\gamma(q^2)$ therefore both vanish at one loop. The same need not be true for the $Z$-boson coupling provided that the $Z$ couplings do change scalar flavour. Just as for fermions, the requirement that the $Z$ vertex change flavour is reflected in the antisymmetry of the final expressions for the scalar-generated anomalous TGV's under the interchange of the masses of the two scalars which connect to the $Z$-boson vertex.

We summarize the one-loop signatures we have found for scalars in column three of Table II.

## 5. Specific Features at One Loop

We next turn to a more explicit evaluation of the one-loop TGV's. We do so in order to determine which parts of the parameter space of the underlying model lead to the largest induced couplings. The parameters that are required can then be compared with other phenomenological bounds to see whether the largest induced TGV's need be ruled out by other phenomenological information.

In order to execute this program, we first compute the dependence of the loop-generated TGV's on the underlying particle masses and couplings. We then use this dependence to determine what values maximize the result. Representative models which can produce these parameters in a phenomenologically acceptable way are then constructed in the next section.

### *5.1) Fermion-Induced TGV's*

To explore the implications of underlying fermion loops for TGV's we evaluate the graph of Fig. (2). For convenience we write the fermion-gauge boson couplings generically as

$$\begin{aligned} \mathcal{L}_{FW} &= ig\overline{\psi}^i\gamma^\mu(A_{ij}\gamma_L + B_{ij}\gamma_R)\psi^j\ W_\mu + \text{h.c.} \\ \text{and} \qquad \mathcal{L}_{FV} &= ie_V\overline{\psi}^i\gamma^\mu(C^V_{ij}\gamma_L + D^V_{ij}\gamma_R)\psi^j\ V_\mu, \end{aligned} \tag{28}$$

where $V$ represents either the $Z$ boson or the photon, and $e_V$ is the corresponding coupling as defined in Section (2). To be completely concrete, in the Standard Model we would have $B_{ij} = 0$ and $A_{ij} = \frac{1}{\sqrt{2}}V_{ij}$, where $V_{ij}$ is the Cabbibo-Kobayashi-Maskawa (CKM) matrix (for quarks). Similarly $C^\gamma_{ij} = D^\gamma_{ij} = Q_{ij}$ for the photon, where $Q$ is the matrix of electric charges chosen such that the proton has charge $+1$. And finally, $C^Z_{ij} = \frac{1}{c_w^2}(T_3 - Qs_w^2)_{ij}$, and $D^Z_{ij} = -\frac{s_w^2}{c_w^2}Q_{ij}$. As usual, $T_3$ here represents the matrix generator of the third component of weak hypercharge.

*The Dipole Moment Form Factor, $f_V$:*

With these convention for the fermion couplings, the result for the induced TGV may be expressed in a compact form in terms of a double integral over Feynman parameters. We find the following contribution to the dipole-moment form factor, $f_V(q^2)$:

$$f_V(q^2) = \frac{g^2}{8\pi^2} \sum_{ij\ell} \Big[ \mathcal{C}_1\, I_1 + \mathcal{C}_2\, I_2 + \mathcal{C}_3\, I_3 + \mathcal{C}_4\, I_4 \Big], \tag{29}$$

where the constants $\mathcal{C}_i$ are given in terms of the fermion couplings by:

$$\begin{aligned}
\mathcal{C}_1 &= -\mathrm{Im}\,(A^*_{i\ell} C^V_{ij} A_{j\ell} - B^*_{i\ell} D^V_{ij} B_{j\ell}), \\
\mathcal{C}_2 &= +\sqrt{\beta_i \beta_\ell}\; \mathrm{Im}\,(A^*_{i\ell} C^V_{ij} B_{j\ell} - B^*_{i\ell} D^V_{ij} A_{j\ell}), \\
\mathcal{C}_3 &= +\sqrt{\beta_j \beta_\ell}\; \mathrm{Im}\,(A^*_{i\ell} D^V_{ij} B_{j\ell} - B^*_{i\ell} C^V_{ij} A_{j\ell}), \\
\text{and} \qquad \mathcal{C}_4 &= +\sqrt{\beta_i \beta_j}\; \mathrm{Im}\,(A^*_{i\ell} D^V_{ij} A_{j\ell} - B^*_{i\ell} C^V_{ij} B_{j\ell}).
\end{aligned} \tag{30}$$

Here $\beta_i \equiv m_i^2/M_W^2$ represent the fermion masses, normalized to the $W$ mass, and the labels '$i$', '$j$' and '$\ell$' are as defined in Fig. (2). The integrals, $I_i$, which arise are given by:

$$\begin{aligned}
I_1 &= \int_0^1 \int_0^1 dx\, dy\; y^2 (1-2x) \left\{ 3 \ln \mathcal{A} - \frac{t}{4} \left[ \frac{1 - y^2 (1-2x)^2}{\mathcal{A}} \right] \right\} \\
I_2 &= \int_0^1 \int_0^1 dx\, dy\; \frac{y\,[1 + y\,(1-2x)]}{\mathcal{A}} \\
I_3 &= \int_0^1 \int_0^1 dx\, dy\; \frac{y\,[1 - y\,(1-2x)]}{\mathcal{A}} \\
I_4 &= \int_0^1 \int_0^1 dx\, dy\; \frac{y^2\,(1-2x)}{\mathcal{A}},
\end{aligned} \tag{31}$$

with the function $\mathcal{A}$ defined by:

$$\mathcal{A} = \beta_\ell (1-y) + \beta_j\, xy + \beta_i (1-x) y - y(1-y) - t y^2 x (1-x). \tag{32}$$

These expressions all depend on $q^2$ through the variable $t \equiv q^2/M_W^2$.

Notice that although $I_1$ and $I_4$ are antisymmetric under the interchange $m_i \leftrightarrow m_j$, $I_2 \leftrightarrow I_3$ under this replacement, so the sum of these terms need not antisymmetric. This reflects the fact that the contributions $I_1$ and $I_4$ rely for their CP-violation on the existence of flavour-changing $Z$ couplings, while $I_2$ and $I_3$ are due to the interference of the left- and right-handed couplings of the $W$. In particular, for the photon the condition $C^\gamma_{ij} = D^\gamma_{ij} = Q_i \delta_{ij}$ implies that $\mathcal{C}_1 = \mathcal{C}_4 = 0$, and eq. (29) reduces to the expression obtained by earlier workers [14] [27].

We have evaluated these integrals numerically, for various values of the fermion masses, $m_i$, $m_j$, $m_\ell$ and of $q^2$. Fig. (7) exhibits an illustrative plot of the real and imaginary parts of $f_V(q^2 = 4M_W^2)$ – where $q^2 = 4M_W^2$ is the value most appropriate for LEP 200 – as a function of $m_i$ with $m_j = 100$ GeV, and $m_\ell = 50$ GeV. These mass values have been chosen in order to display the threshhold enhancement which occurs when $(m_i + m_j)^2 = q^2 = 4M_W^2$.

There are two points concerning this graph that bear emphasis:

• *1:* It is noteworthy that the threshhold enhancement makes the region of relatively small masses give larger contributions than do the large mass regions. This is quite representative of what happens for general combinations of masses. Of course, the presence of a threshhold, and the associated absorptive parts, also show that the new fermions in the loop are themselves being pair produced at these energies. In this case the electroweak form factors are much less interesting than this more spectacular production process, unless the produced fermions should not themselves be detectable at LEP 200. We determine when this can happen in the next Section, where we build explicit models for the underlying physics.

• *2:* For the purposes of the plot we have taken the imaginary parts of the combinations of coupling constants which appear in front of all terms to be one — *i.e.* Im $(\cdots) = 1$ — a fact which must be kept in mind when making contact with specific models. For $CP$-violating TGV's this is overly optimistic to the extent that the $CP$-violating phases of the underlying model should turn out to be small. The same is not true for $CP$-preserving TGV's, however, for which this is not such a restrictive assumption. As may be seen for the Standard-Model couplings given above, the values ultimately taken by these parameters depend upon the representation content of the new particles, as well as on which vector-boson TGV's are under consideration.

*The Anapole Form Factor, $h_V$:*

The graph of Fig. (2) also generates a contribution to $h_Z(q^2)$, which is given by:

$$h_Z(q^2) = \frac{g^2}{8\pi^2} \sum_{ij\ell} \Big[ \mathcal{D}_1 \, I_4 + \mathcal{D}_2 \, I_5 . \Big], \tag{33}$$

Here the constants $\mathcal{D}_i$ are given by:

$$\begin{aligned} \mathcal{D}_1 &= +\sqrt{\beta_i \beta_j}\ \mathrm{Im}\ (A^*_{i\ell} D^Z_{ij} A_{j\ell} + B^*_{i\ell} C^Z_{ij} B_{j\ell}). \\ \mathcal{D}_2 &= +\mathrm{Im}\ (A^*_{i\ell} C^Z_{ij} A_{j\ell} + B^*_{i\ell} D^Z_{ij} B_{j\ell}), \end{aligned} \tag{34}$$

where the labels '$i$', '$j$' and '$\ell$' are as before. $I_4$ is given above in eq. (31), and the new integral, $I_5$, is:

$$I_5 = \int_0^1 \int_0^1 dx\, dy\; y^2(1-2x) \left\{ \ln \mathcal{A} - \frac{t}{4} \left[ \frac{1 - y^2(1-2x)^2}{\mathcal{A}} \right] \right\}. \tag{35}$$

As before, the function $\mathcal{A}$ is defined as in eq. (32). Notice the antisymmetry of both $I_4$ and $I_5$ under the interchange $m_i \leftrightarrow m_j$, as was argued on general grounds in the previous Section.

Evaluating these integrals numerically, for the same values of the fermion masses and of $q^2$ as were used above for $f_V$ — *i.e.* $m_j = 100$ GeV, $m_\ell = 50$ GeV and $q^2 = 4M_W^2$ — leads to the plot of $h_V$ *vs* $m_i$ that is given in Fig. (8). For the purposes this plot, we take the various couplings to satisfy Im $(\cdots) = 1$. Once again we find a threshhold enhancement when $(m_i + m_j)^2 = q^2 = 4M_W^2$ (*i.e.* $m_i = 60$ GeV on the plot), as well as the required zero when $m_i = m_j$ (= 100 GeV on the plot).

### *5.2) Scalar-Induced TGV's*

We next perform for scalars a calculation that is similar to that just presented for fermions. An immediate complication here is the much larger class of graphs which, in principle, must be considered. We therefore restrict our goal here to finding reasonably representative examples with comparatively large TGV's, rather than to exhaustively surveying all models. As a result we compute here only the result for the loop of Fig. (6) (in which only scalars circulate within the loop), as well as the graph of Fig. (5) (in which two scalars circulate within the loop together with a single $W$ boson). In this latter graph we consider only the case where it is the neutral gauge boson ($Z$ or photon) which couples to the scalar line. This class is sufficiently simple to keep the labour involved under control, yet it is also general enough to include most of the phenomenologically interesting models. Previous workers [28] [20] have also restricted their attention to this class.

We parameterize our three-point scalar-vector interactions in the following way:

$$\begin{aligned}
\mathcal{L}_{SSW} &= -igE^{ij}\Big[\phi_i^*\partial^\mu\phi_j - (\partial^\mu\phi_i^*)\phi_j\Big]W_\mu + \text{h.c.} \\
\mathcal{L}_{SSV} &= -ie_V F_V^{ij}\Big[\phi_i^*\partial^\mu\phi_j - (\partial^\mu\phi_i^*)\phi_j\Big]V_\mu \\
\mathcal{L}_{SWW} &= -iG^i M_W\,\phi_i\,W_\mu^* W^\mu .
\end{aligned} \tag{36}$$

For the physical Higgs in the Standard Model these couplings would become: $E = F_\gamma = F_Z = 0$ and $G = 1$.

We may now compute the contributions to $h_V(q^2)$ from the Feynman graphs of Figs. (5) and (6). As discussed in the previous section, both graphs give zero for the electromagnetic form factors, $h_\gamma(q^2)$, and $\hat{h}_\gamma(q^2)$. For $V = Z$, the purely scalar loop of Fig. (6) gives:

$$(h_V)_{\text{Fig.(6)}} = \frac{g^2}{8\pi^2}\sum_{ij\ell} \text{Im}\,[(E^{i\ell})^* F_Z^{ij} E^{j\ell}]\; J_1, \tag{37}$$

while evaluation of the graph of Fig. (5) (with the $Z$ coupled to the scalar line) gives

$$(h_V)_{\text{Fig.(5)}} = -\frac{g^2}{8\pi^2}\sum_{ij\ell} \text{Im}\,[(G^i)^* F_Z^{ij} G^j]\; J_2. \tag{38}$$

In these expressions $J_i$ represent the following two-dimensional Feynman parameter integrals:

$$\begin{aligned}
J_1 &= \int_0^1\int_0^1 dx\,dy\; y^2(1-2x)\Bigg\{16\ln\mathcal{A} \\
&\qquad + \frac{1}{6}\left[\frac{t[16y^2x(x-1) - 3(1+2y)] + 4(1+2y)(1+y)}{\mathcal{A}}\right]\Bigg\} \\
J_2 &= \int_0^1\int_0^1 dx\,dy\; y^2(1-2x)\left\{\frac{8}{3}\ln\mathcal{A} + \frac{2}{3}\left[\frac{ty^2x(1-x) + \beta_\ell - y(1-y)}{\mathcal{A}}\right]\right\}
\end{aligned} \tag{39}$$

The quantities $\beta_i$ again represent the mass ratios $m_i^2/M_W^2$, $t$ is again $q^2/M_W^2$, and $\mathcal{A}$ represents the same function of these quantities as was given in eq. (32).

Some explanation is in order as to the meaning of the labels '$i$', '$j$ and '$\ell$'. For Fig. (6), these labels refer to the species of scalars circulating in the loop, as defined in the figure. For Fig. (5), however, although '$i$' and '$j$' are also defined in this way, '$\ell$' runs over the species of vector particle inside the loop. Typically this internal gauge boson is the $W$ particle, in which case $\ell = W$ and $\beta_\ell = 1$. Notice that both of the integrals $J_1$ and $J_2$ have the expected antisymmetry under the interchange of the two scalar masses: $m_i \leftrightarrow m_j$.

In Figs. (9) and (10) we present a representative plot of these expressions as functions of their mass arguments. For the purposes of comparison with the fermion results presented earlier, in Fig. (9) we use the same masses as were used in Figs. (7) and (8): $q^2 = 4M_W^2$, $m_j = 100$ GeV, and $m_\ell = 50$ GeV. For Fig. (10) we use the same values for $m_i$ and $m_j$, but use $m_\ell = M_W$. For the purposes of these plots we also take the combination of couplings which premultiplies the integrals, Im $(\cdots)$, to equal one.

Once again the threshhold enhancement is visible for $m_i = \sqrt{q^2} - m_j = 60$ GeV. And once again, this indicates the direct pair production of the underlying scalars, which may or may not be detectable depending on their properties.

*5.3) The Heavy-Mass Limit*

It is instructive to compare the above explicit expressions with the general form that is dictated by the effective lagrangian analysis of Section (3), in the limit that the masses of the particles in the loops are large. In this limit our general expression must reduce to the effective-lagrangian result. We make the following three remarks.

• *(1):* The most basic prediction of the effective lagrangian is that the form factors are approximately independent of $q^2$, being well described by their values at $q^2 = 0$. It is clear, as we have verified numerically, that this $q^2$-independent behaviour may be expected to hold up to powers of $q^2/M^2$ for any $q^2 \ll M^2$, where $M$ is the underlying large particle mass. Since, for LEP-200 energies, $q^2 = 4M_W^2$, $q^2/M^2$ is as small as 10% for $M \gtrsim 500$ GeV.

• *(2):* A second, more model-dependent, prediction of an effective theory concerns how quickly the results fall off as the heavy mass gets large. We next compare our explicit one-loop calculation with the estimates of Table I.

Consider separately two cases according to whether or not the new $CP$-violating physics also breaks the electroweak gauge group. The simplest situation to analyse is that for which the new physics is $SU_L(2) \times U_Y(1)$-invariant. In this case it is generally possible to make the scale of new physics arbitrarily large in comparison to $v$. The LRDA estimate then predicts that all anomalous TGV's should fall off at least as quickly as $1/M^2$.

For the one-loop calculation, an $SU_L(2) \times U_Y(1)$-invariant way of taking the heavy-mass limit is to take the mass, $M$, of the heavy scalar or fermion multiplet to be large, while taking no mass-splittings within this multiplet, $\Delta M \to 0$. With this choice our one-loop results indeed reproduce the predicted $1/M^2$ behaviour. For scalars the loop result simply vanishes for $\Delta M = 0$, as do the contributions of the integrals $I_1$ and $I_4$ to the fermion loop. The fermion contribution $\mathcal{C}_2 I_2 + \mathcal{C}_3 I_3$, however, is nonzero in this limit, and it indeed falls off as $M_W^2/M^2$.

The large-$M$ limit is more delicate when the new physics is not $SU_L(2) \times U_Y(1)$ invariant, because in this case we cannot take $M$ to be larger than $\sim 4\pi v$. As a result factors of $v/M$ can be numerically difficult to separate from factors of $1/4\pi$. Fig. (11) illustrates the result of the loop calculation, where we plot $f_V(q^2 = 4M_W^2)$ (which is real) against $M$, where the particle masses are chosen as $m_\ell = m_j/2 = m_i/3 \equiv M$. (Again we take $\mathcal{C}_i = 1$.) As may be seen from the figure, the result quickly flattens off, becoming essentially $M$-independent, as $M$ grows. Intuitively, this behaviour arises because the dominant contributions – in this example $I_1$ dominates – are proportional to a power of $\Delta M/M$, which is fixed given our choices for particle masses. (A similar result holds for the scalar loop.)

At first sight this behaviour seems to contradict the $M$-dependence of the NDA estimate of Table I, since this indicates a $1/M^2$ dependence. This contradiction is illusory, though, since Table I does not keep track of any of the dimensionless constants of the underlying calculation. In particular it misses the fact that the coupling between the heavy fermion (or scalar) and the longitudinal gauge boson is given by a Yukawa (or scalar self-) coupling: $\lambda \sim M/v \sim gM/M_W$. Since we vary $M$ with both $g$ and $M_W$ fixed, this coupling grows with $M$. Including the loop factor $1/16\pi^2$, gives a more precise version of the NDA estimate, which includes more of the details of the underlying model:

$$\begin{aligned} \tilde{\kappa}_V &\sim \left(\frac{\lambda}{4\pi}\right)^2 \left(\frac{M_W}{M}\right)^2 \\ &\sim \left(\frac{g}{4\pi}\right)^2 . \end{aligned} \qquad (40)$$

We use the relation $\lambda \sim gM/M_W$ in getting the last line. This refined estimate implies an $M$-independent result whose size is $O(10^{-3})$. This agrees well with both the behaviour and the size of the explicit loop result. Notice that this is often smaller than the largest value that is obtained for light fermions (and scalars) using the threshhold enhancement.

• *(3):* The final feature which we explore is the nature by which the effects of the custodial symmetry arise in the specific models. Within the underlying theory, just as for the Standard Model, there are two types of interactions which break the custodial symmetry. They are (*i*) the $U_Y(1)$ gauge couplings, and (*ii*) mass splittings between the members of an $SU_L(2)$ multiplet. The custodial limit therefore corresponds to the situation where all components of the multiplet are degenerate in mass, and for which the matter couplings to the $U_Y(1)$ gauge potential, $B_\mu$, are turned off.

Consider separately the contributions to the $P$- and $CP$-violating moments, $f_V(0)$ (or $\tilde{\kappa}_V$) and the $C$- and $CP$-violating moments $h_V(0)$ (or $a_V$). For $f_V$, only the fermion loop need be considered, since only fermions can contribute to this particular form factor. We have seen that the custodial limit of the effective lagrangian permits both $\tilde{\kappa}_Z$ and $\tilde{\kappa}_\gamma$ to be nonzero, provided that they are related by $\tilde{\kappa}_Z = -\tilde{\kappa}_\gamma \, s_w^2/c_w^2$. In the explicit calculation a nonzero result for these couplings is indeed obtained. Since the couplings to $B_\mu$ vanish in the custodial limit, the photon and $Z$-boson TGV's are simply related in this limit by the relative size of their overlap with $W_\mu^3$. Keeping in mind the difference in normalizations — *i.e.*: $e_\gamma = e$ while $e_Z = ec_w/s_w$ — we indeed find the required relation between $\tilde{\kappa}_\gamma$ and $\tilde{\kappa}_Z$ in the custodial limit.

For the anapole form factors, $h_V(0)$ and $\hat{h}_V(0)$, the argument proceeds differently. Inspection of Table I indicates that, within the effective lagrangian, unbroken custodial symmetry implies that the $Z$-moments, $h_Z(0)$ and $\hat{h}_Z(0)$, must vanish. This is indeed the result of the underlying calculation in the custodial limit, since in this limit all of the internal scalars must be degenerate in mass. But we've seen that this moment vanishes if $m_i = m_j$, regardless of whether it is produced by underlying fermions or by scalars. As a result it must also vanish in the limit when both $m_i$ and $m_j$ are taken to zero, as required.

## 6. Model Building

Our final goal is to construct illustrative models which can generate sizable TGV's and yet satisfy all current bounds. We also ask that the light particles which circulate within our loops not be themselves directly detectable at LEP 200. Our motivation for doing so is to search for an 'existence proof' that anomalous TGV's of the size we have found might be the first indications of new physics. In so doing we consider separately the cases of new heavy fermions and scalars.

Although we can construct an acceptable model of underlying scalar physics, we are unable to do so for fermions. The reason for this is tied to the fact that we are working only with $CP$-violating anomalous TGV's, for which light fermion electric dipole moments furnish extremely strong constraints. We do not expect to encounter the same obstacles for fermion models which produce sizable $CP$-conserving form factors. Unfortunately, however, the search for any such models lies outside the scope of our calculations. These conclusions are described in the present section.

The constraint which drives the search for models is the following question: How can a 'phenomenologically interesting' – *i.e.* few % – anomalous $WWZ$ coupling be generated without also creating a phenomenologically disasterous $WW\gamma$ coupling of similar size?[7]

*6.1) A Fermion Model*

The above question has foiled our attempts to construct a viable model using underlying fermions. To see why this is so, consider what the model must satisfy. It must: $(a)$ contain fermions in the (50 – 100) GeV mass range in order to take advantage of the threshhold enhancement; $(b)$ violate $CP$; $(c)$ couple strongly enough to the $Z$ boson to provide a sizable $f_Z(q^2)$; and yet $(d)$ not couple strongly enough to the photon to induce a similarly large $f_\gamma(q^2)$. In order to avoid having TGV's be irrelevant, we would also ask that the pair-production of these fermions not itself be directly detectable at LEP.

Two methods suggest themselves for suppressing the electromagnetic TGV's without also removing those for the $Z$. Either: add electroweak multiplets and suppress their contribution to $f_\gamma(q^2)$ by arranging for their electrically-charged members to become very massive; or add electroweak singlets, and have these develop $Z$ couplings by mixing with the SM neutrinos. We argue here that neither of these approaches leads to models which satisfy properties $(a)$ through $(d)$.

We consider the second option first: introduce only electroweak singlets, and have them develop $Z$ couplings through mixing with SM neutrinos. The difficulty with this scenario is that the mixing can never be strong enough to generate a sufficiently large $WWZ$ TGV. Recall that the mixing angle, $\theta$, between a heavy electroweak-singlet state, $S$, of mass $M$, and a SM neutrino state, of mass $m$, is determined by diagonalizing the fermion mass matrix. This diagonalization gives a mixing angle that is (in order of magnitude) given by $\theta \sim \sqrt{m/M}$. Since we require $M \gtrsim 50$ GeV to being already ruled out by experiments at the $Z$ resonance, and since even the tau-neutrino mass is bounded to be smaller than $m \lesssim 30$ MeV, we see that $\theta \lesssim 10^{-2}$. Since the induced $WS\tau$ and $ZS\nu_\tau$ couplings are $O(\theta)$ in size, we see that the TGV-inducing loop must be suppressed by at least $\theta^2 \lesssim 10^{-4}$. It therefore produces a $WWZ$ vertex which is many orders of magnitude too small to be detected.

In the alternative scenario we add a nontrivial electroweak multiplet, and attempt to suppress the contribution of its electrically-charged states to the TGV loop by making these states very heavy. At first glance this appears to be possible, since all of the loops which contribute to $f_\gamma$ involve at least two heavy charged-particle propagators, while $f_Z$ can be generated from graphs with only one such propagator. (Recall that the photon must be attached to the heavy charged-particle line.) Unfortunately, this observation does not parlay itself into a suppression of $f_\gamma$ relative to $f_Z$. This is because the graphs involving only one heavy-particle propagator, which are supposed to produce the large contribution to $f_Z$, must actually vanish in the limit where the neutral-particle masses, $M_n$, are much smaller than the charged particle masses, $M_c$. After all, in this limit the two fermions which couple to the $Z$ – that is to say: the neutral particles – are effectively degenerate: $m_i \sim m_j \sim M_n \ll M_c$. And we've seen in previous sections that all of the contributions to $f_V$ which are not explicitly proportional to $m_i$ or $m_j$ vanish when $m_i = m_j$. As a result, the graphs which contribute differently to $f_Z$ and $f_\gamma$ actually turn out to be suppressed by a power of $M_n/M_c$. The same is not true of the graphs that $f_Z$ and $f_\gamma$ have in common. The consequence is that raising $M_c$ relative to $M_n$ does *not* lead to a suppression of $f_\gamma$ relative to $f_Z$.

[7] Such a couping is only a disaster for the $CP$-violating TGV's, for which the bounds from the neutron and electron electric dipole moments apply.

There is yet another difficulty with this approach. Notice that the limit $M_c \gg M_n \sim M_W$ strongly breaks the custodial symmetry which protects the prediction $\Delta\rho \lesssim O(1\%)$. With this pattern of masses, vacuum-polarization loops involving the heavy multiplet typically contribute an unacceptable amount to $\Delta\rho$. If $\Delta M^2 \sim M_c^2 - M_n^2 \gg M_W^2$ denotes the size of the mass splitting within the new multiplet, then $\Delta\rho \sim (\alpha \Delta M^2/4\pi M_W^2)$. So we would expect that even if a large charged-fermion mass could enhance $f_\gamma$, the custodial-symmetry breaking in any such model would lead to an unacceptably large size for $\Delta\rho$.

### *6.2) A Scalar Model*

There is much more latitude in building a model of underlying scalars, since in this case there is no need to suppress the electromagnetic TGV. This is because scalars can only generate $h_V$ and $\hat{h}_V$, and these are poorly constrained for *both* $V = Z$ and $V = \gamma$. In this section we construct a viable scalar model which generates maximal TGV's, as an existence proof that such models are possible.

The model consists of supplementing the Standard Model Higgs, $\phi$, with two additional scalar multiplets, $\chi^\alpha$ and $S$. We choose $\chi^\alpha$ to be a real $SU_L(2)$ triplet with vanishing hypercharge, $Y_\chi = 0$, and we take $S$ to be a complex singlet with $Y_S = -1$.[8] Two multiplets are added in order to allow $CP$ invariance to be broken by the renormalizable potential, as well as to permit off-diagonal $Z$-scalar couplings. The slightly unorthodox quantum numbers have a useful spinoff in that they do not permit the dimension-four fermion-flavour-changing Yukawa couplings to charged fermions, which can be a problem for multi-doublet models.

[8] Our hypercharge conventions are such that the SM Higgs has $Y_\phi = \frac{1}{2}$.

We require vanishing *v.e.v.*s for both of the new fields. In order to prevent $\chi_\alpha$ from developing an expectation once $\phi$ does, as would be possible through a $\phi\phi\chi$ term in the potential, we impose a discrete symmetry under which the new fields are simultaneously reflected, $(\chi, S) \to (-\chi, -S)$, while all SM fields are invariant. The most general renormalizable scalar potential that is possible for such a model is

$$\begin{aligned} V(\phi, \chi, S) = \lambda_\phi \left(\phi^\dagger\phi - \frac{v^2}{2}\right)^2 + \frac{1}{2}\mu_\chi^2\, \chi^\alpha\chi_\alpha + \mu_S^2\, S^*S + \lambda_\chi\, (\chi^\alpha\chi_\alpha)^2 + \lambda_S\, (S^*S)^2 \\ + \lambda_{\phi\chi}\, \phi^\dagger\phi\, \chi_\alpha\chi^\alpha + \lambda_{\phi S}\, \phi^\dagger\phi\, S^*S + \lambda_{\chi S}\, \chi^\alpha\chi_\alpha\, S^*S \\ + \left[\lambda_{\phi\chi S}\, (\tilde{\phi}^\dagger t_\alpha\phi)\, \chi^\alpha S + \text{c.c.}\right]. \end{aligned} \tag{41}$$

Here the $t_\alpha$ are generators of $SU_L(2)$, and $\phi$ is the usual SM Higgs. $\tilde{\phi} = -it_2\phi^*$ is the conjugate of the SM doublet. The reality of the lagrangian implies that all of the parameters in $V(\phi, \chi, S)$ must be real, with the exception of $\lambda_{\phi\chi S}$, which is the model's source of $CP$-violation.

The parameters of this potential are chosen to ensure that $\langle\chi_a\rangle = \langle S\rangle = 0$, and $\langle\phi\rangle = \frac{1}{\sqrt{2}}\binom{0}{v}$. Expanding about this solution, we see that the scalar spectrum consists of two electrically-charged and two neutral scalars. The neutral-scalar mass eigenstates consist of the usual Higgs, having mass $m_H^2 = 2\lambda_\phi v^2$, as well as $\chi_3$, which has mass $m_3^2 = \mu_\chi^2 + \frac{1}{2}\lambda_{\phi\chi}v^2$. The charged states, $S$ and $\chi = \frac{1}{\sqrt{2}}(\chi_1 + i\chi_2)$, mix through the following complex hermitian mass matrix:

$$-\frac{1}{2}\begin{pmatrix} S \\ \chi \end{pmatrix}^\dagger \begin{pmatrix} \mu_S^2 + \frac{1}{2}\lambda_{\phi S}v^2 & -\frac{1}{4}\lambda_{\phi\chi S}^* v^2 \\ -\frac{1}{4}\lambda_{\phi\chi S}v^2 & \mu_\chi^2 + \frac{1}{2}\lambda_{\phi\chi}v^2 \end{pmatrix} \begin{pmatrix} S \\ \chi \end{pmatrix}, \tag{42}$$

which may be diagonalized by performing a two-by-two unitary rotation amongst the two charged scalars. This rotation involves a $CP$-violating phase provided that $\lambda_{\phi\chi S}$ is complex.

We may now compute the loop-induced anomalous TGV within this model. The first step is to determine which graphs can potentially be nonzero. Since our new fields have vanishing *v.e.v.*s, there are no $\chi VV'$ or $SVV'$ vertices in this model, where $V$ and $V'$ represent either $W$ or $Z$. This rules out any contribution from the graphs of Figs. (3) and (5). At first glance, those of Fig. (4) could contribute since the model has $W^*W\chi^*\chi$, $W^*WS^*S$, $W^*W\chi_3^2$ and $W^*Z\chi\chi_3$ vertices. None of these graphs contributes in the end, however. The $W^*W$-type vertices can never contribute because they do not have the correct tensor structure to provide a contribution to $h_V$. This same objection does not apply to the $W^*Z\chi\chi_3$ vertex or its hermitian conjugate, although these nevertheless give no contribution to $CP$-violating TGV's, since the result vanishes once the contribution from all possible internal scalars is summed. In the end, the only possible nonzero graph is that which involves just scalars circulating around the loop, giving the result proportional to $J_1$ in eq. (33).

We are free to choose the scalar masses in the $(60-100)$ GeV range which is consistent with the threshhold enhancement of the TGV integral. This is also large enough to evade direct production bounds at the $Z$ resonance. We must also consider their indirect effects, which arise through their loop contributions to precision electroweak measurements. Since the new scalars have no Yukawa couplings to any of the charged SM fermions, their loop effects are completely described by their contribution to the $W$ and $Z$ vacuum polarizations – the 'oblique' corrections. Because the intended masses of our scalars are not large compared to $M_W$, these oblique corrections cannot be parameterized in terms of the original variables, $S$, $T$ and $U$, of Ref. [10]. We must instead adopt the six-parameter generalization of Ref. [11], which *is* applicable to such comparatively light particles.

Computing the scalar contributions to the weak-boson vacuum polarizations [30], and comparing the results to the global fit of Ref. [11], we see that our two scalar multiplets contribute acceptably to oblique corrections. An intuitive understanding of this follows from the observation that each real scalar contributes roughly $1/2\pi \simeq 0.16$ to the oblique parameters, such as $S$. So the contribution from our five real scalars is roughly five times as large: $S \simeq 0.8$. But the 2-$\sigma$ bound for $S$ that is found in Ref. [11] is $-4.3 < S < 2.5$, with similar bounds holding for the other parameters. Our five real scalar states therefore slip inside these bounds at the present level of experimental precision.

It is noteworthy that a fit involving the same data, but using only the variables $S$, $T$ and $U$, gives the $2-\sigma$ limit of $-1.3 < S < 0.3$ [11], which would rule out the model we have presented. This is because, to the extent that the entire parameter space of the oblique corrections can be accessed within the underlying theory, the more parameters that contribute to the fit, the wider is the allowed range. As a result the bound for the light scalars is actually *weaker* in these circumstances than is the bound that would be expected for TeV scalars.

We finally ask whether the pair-production of these scalars would be itself directly detectable at LEP 200. In this model there is a conserved quantum number associated with a common overall rotation of only the new scalar fields. As a result, the lightest exotic scalar is stable, and the heavier scalars can only decay into a gauge boson plus a lighter scalar which would escape detection. If this reaction is energetically possible, then above the $W$ threshhold, the signal for the production of these new scalars would be spectacular: a pair of $W$ bosons which are not necessarily back-to-back, with appreciable missing energy. The decay would be impossible, however, if the mass splitting amongst the scalars is too small to permit the decay $\phi \to W\,\phi'$. In this case all of the scalars escape detection, and they could be expected to be first found through their contribution to the gauge boson anomalous TGV's.

## 7. Conclusions

In this paper we report on a detailed study of the kinds of CP-violating, three-point electroweak-boson self-couplings that are generated at one loop by generic new physics involving particles of spins zero through one. We identify several of the generic implications of this type of new physics, and illustrate these general properties with a few more detailed illustrative models.

Several general conclusions emerge from this study. We consider two situations for which general remarks can be made: the case where the new physics is much heavier than the weak scale, and the situation where we make no assumptions concerning the new-particle masses, but for which the contributions to TGV's arise at one loop. Each of these two scenarios has its own distinguishing features, which might be used to diagnose the origins of anomalous TGV's, should these ever be observed. We summarize these conclusions for each of the two scenarios in the following paragraphs.

### *7.1) Heavy New Physics*

Should the new physics be much heavier than the weak scale, then its effects may be analyzed in terms of the low-energy effective lagrangian which would be obtained by integrating out the heavy particles. The utility of such a lagrangian at LEP energies relies on the ability to expand observables in powers of $M_W/M$, where $M$ is the new-physics scale, and so its application requires $M \gg M_W$. In this limit we find the following general features (which apply equally well to both $CP$-violating and $CP$-preserving TGV's):

• *(1):* Although some anomalous TGV's arise at dimension four in the effective lagrangian, they are generically expected to be suppressed by powers of $M_W/M$ in the two most plausible estimates for their sizes. These estimates are necessarily somewhat model-dependent, however, and we consider here for comparison the two examples of Naive Dimensional Analysis and Linearly-Realized Dimensional Analysis. NDA encodes power-counting arguments that work well for strongly-coupled theories like QCD, while LRDA simply consists of ordinary dimensional analysis using the lagrangian which linearly realizes the electroweak gauge group, $SU_L(2) \times U_Y(1)$. The estimates we obtain by using these two schemes are displayed in Table I.

We find that although these two kinds of estimates are based on different physical assumptions concerning the underlying theory, they agree for the strength of many, although not all, of the effective gauge-boson self-interactions. In particular they differ on their predictions for the couplings $\tilde{\lambda}_V$, which are expected to be $O(g^2 v^2/M^4)$ using NDA, but could be $O(g^2/M^2)$ in the linearly-realized theory. Finally LRDA implies $\tilde{\lambda}_Z = \tilde{\lambda}_\gamma$, up to higher-dimension, $O(g^2 v^2/M^4)$, corrections.

• *(2):* We find that a custodial symmetry, which keeps the $\rho$ parameter near unity, enforces several relations among $CP$-violating TGV's. The custodial limit implies $\tilde{\kappa}_Z$ and $\tilde{\kappa}_\gamma$ are related by $\tilde{\kappa}_Z = -\tilde{\kappa}_\gamma \, s_w^2/c_w^2$. The couplings $a_Z$ and $\hat{a}_Z$ must also vanish in this limit. *No conditions at all are implied for* $\tilde{\lambda}_Z$ *or* $\tilde{\lambda}_\gamma$*.*

• *(3):* Should the new physics not break the electroweak gauge group, and if this symmetry is broken by $SU_L(2)$-doublet scalars, then the generic prediction $\tilde{\lambda}_Z = \tilde{\lambda}_\gamma$ follows. Although this latter relation is sometimes claimed in the literature to follow from custodial symmetry, we see here that this is incorrect. It is instead a robust feature of the linearly-realized scenario, making it a useful diagnostic of the nature of the underlying physics should anomalous TGV's be detected.

*7.2) One-Loop TGV's*

Another situation for which general conclusions are possible is when the new physics which produces the TGV's is perturbative. In this case the type of $CP$-violating form factors that are generated directly reflects the nature of the relevant degrees of freedom of the underlying physics. In contrast with the effective-lagrangian analysis, the conclusions we draw in this case are more specific to $CP$-violating TGV's. We find that there are two kinds of couplings which can lead to $CP$-violating TGV's at one loop. These are either spin-half fermion/gauge-boson interactions, or scalar/gauge-boson interactions. Each of these produces a distinctive signature for the anomalous TGV's, which are summarized in Table II. We find:

• *(1):* If the new particles whose $CP$-violating couplings generate the TGV's are spin-half fermions, then we expect the heirarchy $f_V, h_V \gg g_V$. This is because only $f_V$ and $h_V$ can be generated at one loop, while the others cannot receive contributions at less than two loops. $g_V$ does not arise at one loop simply because at this order the required tensor structure does not appear. The only situation for which this heirarchy between the sizes of $f_V$ and the other TGV's need not hold is when one-loop contributions to $f_V$ turn out to themselves be zero.

• *(2):* If scalars are responsible for $CP$-violating TGV's, then we expect $h_V, \hat{h}_V \gg f_V, g_V$, since the $CP$-violating vertices which can contribute at one loop always also break $C$. This ensures that neither $f_V$ nor $g_V$ can be generated at less than two loops. Unless the one-loop contributions happen to $h_V$ and $\hat{h}_V$ happen to vanish, these form factors are therefore expected to be the largest.

*7.3) A Promising Model*

We have searched the parameter space of the underlying model to determine what kinds of physics can maximize the TGV's that they produce. We find that the largest TGV's are generated by particles whose masses are of order $M_W$. In this case, if two of the particles masses satisfy $m_i + m_j = \sqrt{q^2}$ ($= 2M_W$ at the $W$ pair-production threshhold), it is possible to have a threshhold enhancement of the one-loop graph, giving TGV's that can be several percent in size.

Notice that these largest TGV's need *not* generated by physics at the TeV scale, contrary to the intuition that would be based on simple coupling-constant counting. After all, it is heaviest particles which couple most strongly to the longitudinal components of the $W$ and the $Z$, with strength $\sim gM/M_W$. According to this observation, the maximum TGV should be produced by the heaviest virtual particles. In practice, however, we find that this coupling growth with particle mass is compensated by a dependence of the loop on $1/M^2$, leaving a result which is roughly independent of $M$ for large $M$.

We finally look for specific models which can have parameters which can produce the maximal TGV's, and yet not conflict with other phenomenological information. We can do so if the underlying $CP$-violation comes from the coupling of scalar particles. In this case particles can exist in the required mass range with the required couplings without having been hitherto detected, either in direct production experiments, or in precision electroweak measurements. They can also be not directly detectable at LEP 200 itself, even though their contributions to TGV's may be enhanced by threshhold effects.

For underlying fermions, however, we are unable to build an phenomenologically acceptable model. This is because, for fermions, a detectably large contribution to $f_Z$ is always accompanied by a contribution to $f_\gamma$ that is of similar size. Such a large electromagnetic form factor would generate an electron and neutron electric dipole moment that is several orders of magnitude larger than the current bound. This same objection would not apply to the $CP$-invariant TGV's, which might potentially receive large contributions from weak-scale fermions.

We conclude that anomalous TGV's that are as large as 1% or so are not impossible, and should be searched for. Their detection would certainly point to new physics and, if detected in $CP$-violating TGV's, most likely to the existence of relatively light new scalars.

### Acknowledgments

We would like to acknowledge many conversations with David London on the subject of custodial symmetries and TGV's, as well as Elie Boridy, Gilles Couture and Mourad Essid for lending a hand on several key occasions. This research was partially funded by N.S.E.R.C. of Canada, les Fonds F.C.A.R. du Québec, and the Swiss National Foundation.

## Figure Captions

- *Figure (1):* The Feynman graphs through which gauge-boson self-interactions might (but don't) contribute to anomalous $CP$-violating $WWV$ vertices.

- *Figure (2):* The Feynman graphs through which fermion/gauge-boson couplings contribute to anomalous $CP$-violating $WWV$ vertices.

- *Figure (3):* A class of Feynman graphs through which scalar-vector-vector and gauge-boson self-interactions might contribute to anomalous $CP$-violating $WWV$ vertices. This class is meant to include all possible graphs for which one scalar and two vector particles circulate in the loop.

- *Figure (4):* A class of Feynman graphs through which scalar-scalar-vector-vector interactions might contribute to anomalous $CP$-violating $WWV$ vertices.

- *Figure (5):* A class of Feynman graphs in which a loop containing two scalar particles and one vector one can contribute to anomalous $CP$-violating $WWV$ vertices. In most applications only the graphs for which the external $Z$ couples to two different types of scalars (labelled $i$ and $j$) violate $CP$. There are two types of such graphs according to whether or not these scalars are electrically charged or neutral.

- *Figure (6):* The Feynman graphs contributing to anomalous $CP$-violating $WWV$ vertices which come from loops which involve only scalar particles.

- *Figure (7):* This figure plots the contribution to $f_V(q^2 = 4M_W^2)$ of a loop of fermions – Fig. (2) – whose masses are $m_j = 100$ GeV and $m_\ell = 50$ GeV. The result is plotted against $m \equiv m_i$.

- *Figure (8):* This figure plots the contribution to $h_V(q^2 = 4M_W^2)$ due to the fermion loop of Fig. (2), with masses chosen as $m_j = 100$ GeV and $m_\ell = 50$ GeV. The result is plotted against $m \equiv m_i$.

• *Figure (9):* Here we plot the contribution to $h_V(q^2 = 4M_W^2)$ due to a purely scalar loop with masses chosen to be $m_j = 100$ GeV and $m_\ell = 50$. The result is plotted against $m_i$.

• *Figure (10):* Here we plot the contribution to $h_V(q^2 = 4M_W^2)$ due to a scalar/gauge-boson loop. We take one of the scalar masses to be $mj = 100$ GeV and plot the result against the other scalar mass, $m_i$. The vector-boson mass is chosen to be $m_\ell = M_W$.

• *Figure (11):* Here we plot the contribution to $f_V(q^2 = 4M_W^2)$ due to a fermion loop for which all of the masses can become large. We choose $m_\ell = m_j/2 = m_i/3 = m$, and plot the result against $m$.

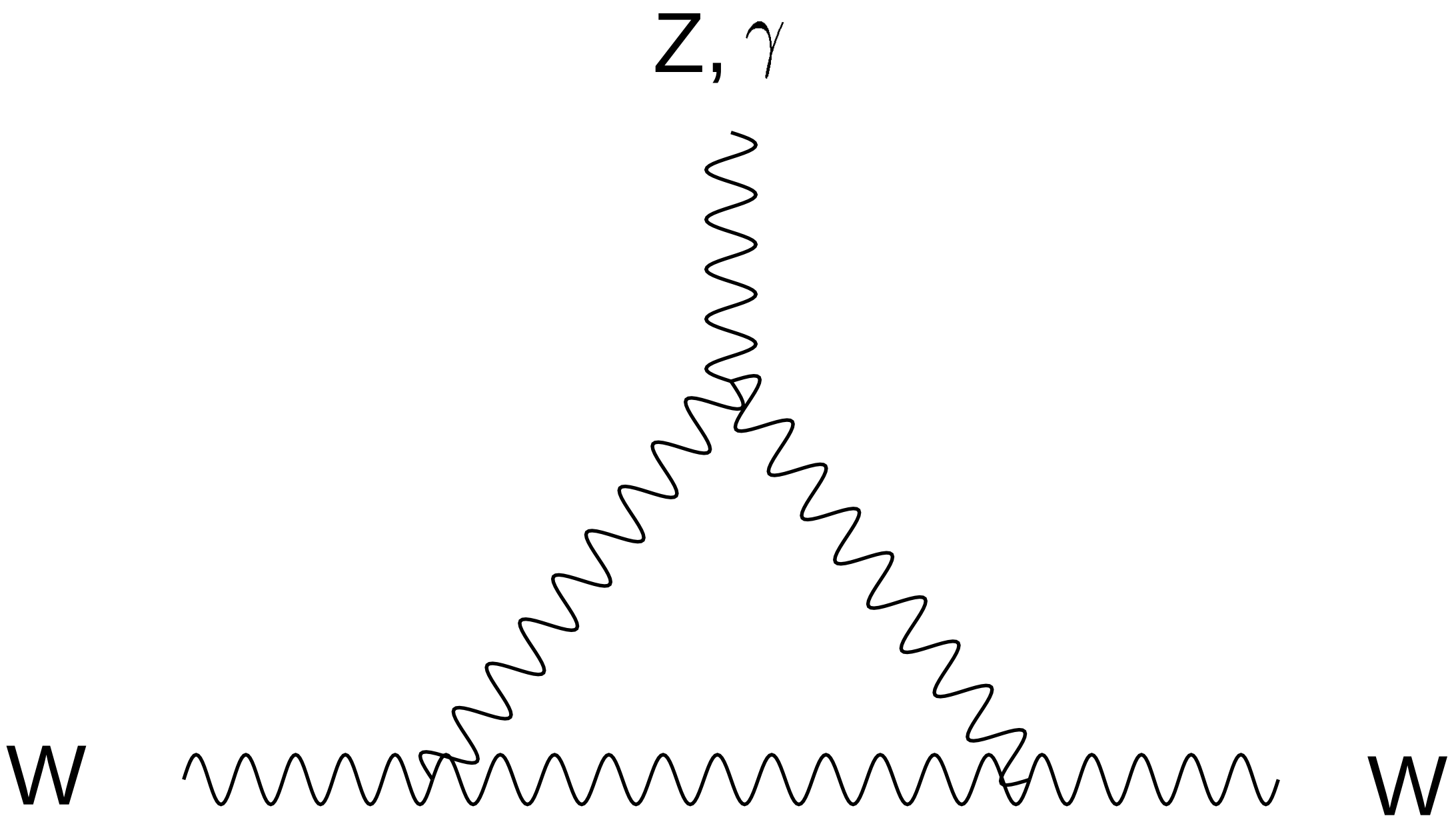

Figure 1

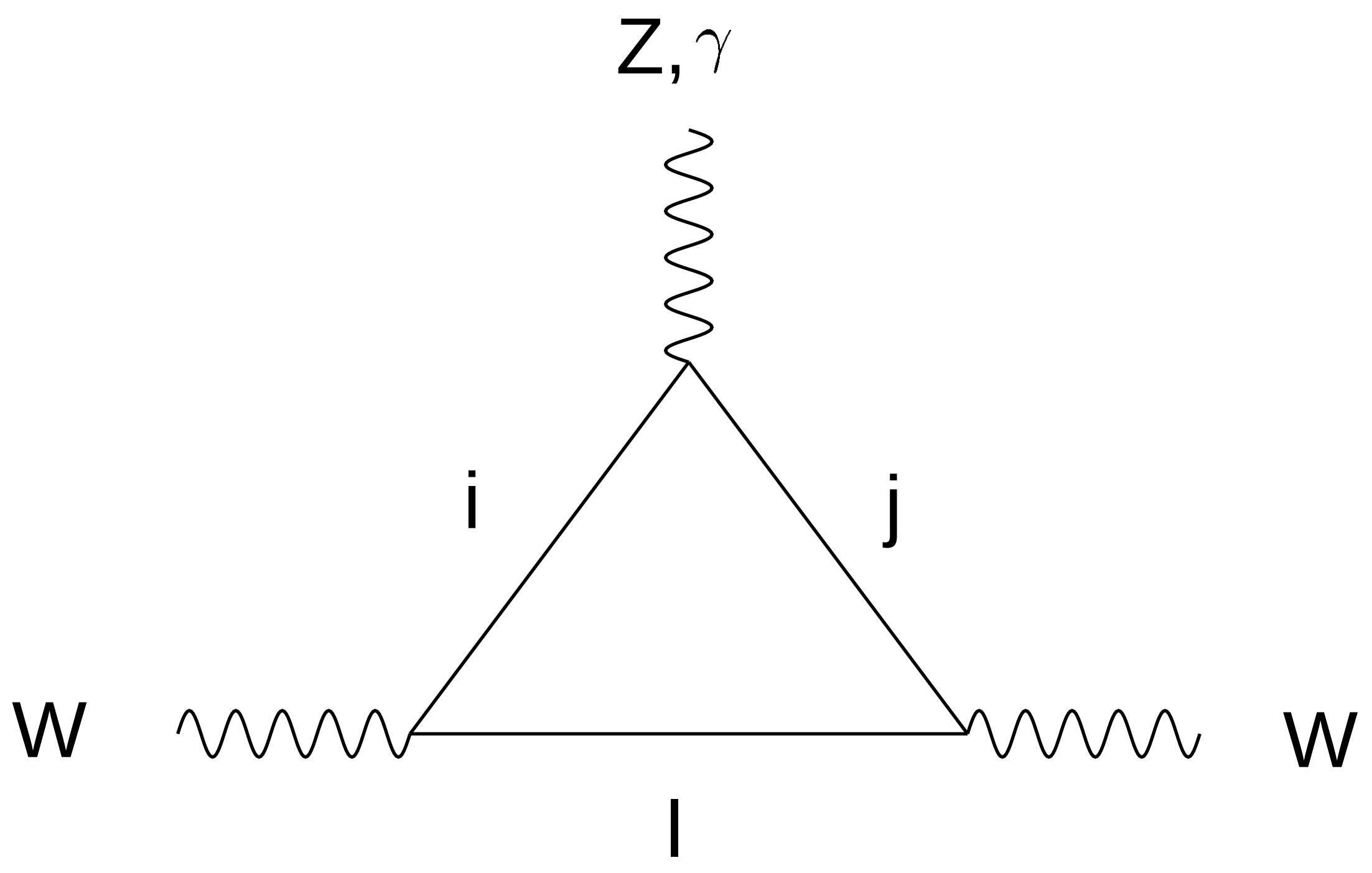

Figure 2

This figure "fig3-2.png" is available in "png" format from:

http://arxiv.org/ps/hep-ph/9312238v2

This figure "fig3-3.png" is available in "png" format from:

http://arxiv.org/ps/hep-ph/9312238v2

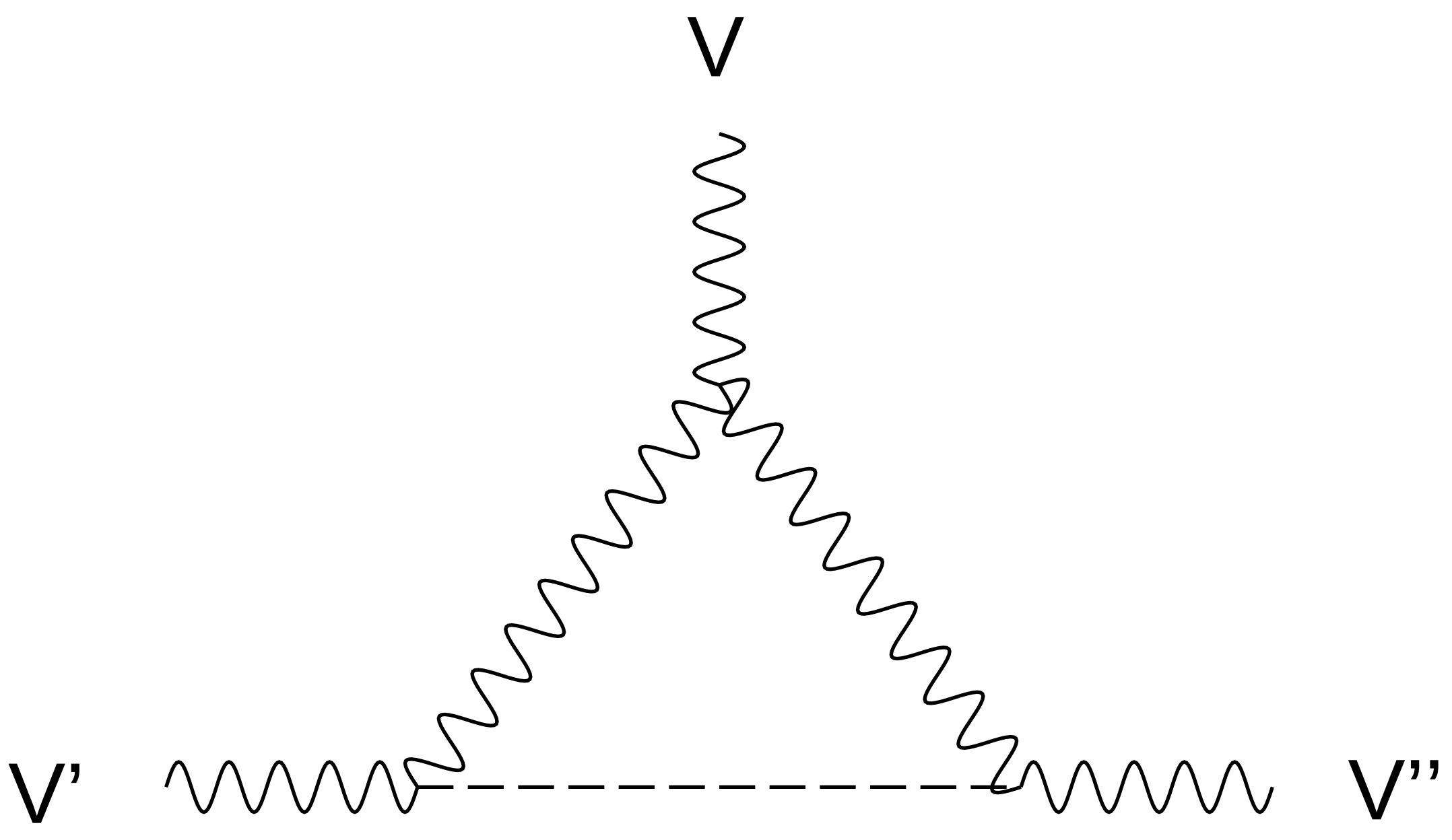

Figure 3

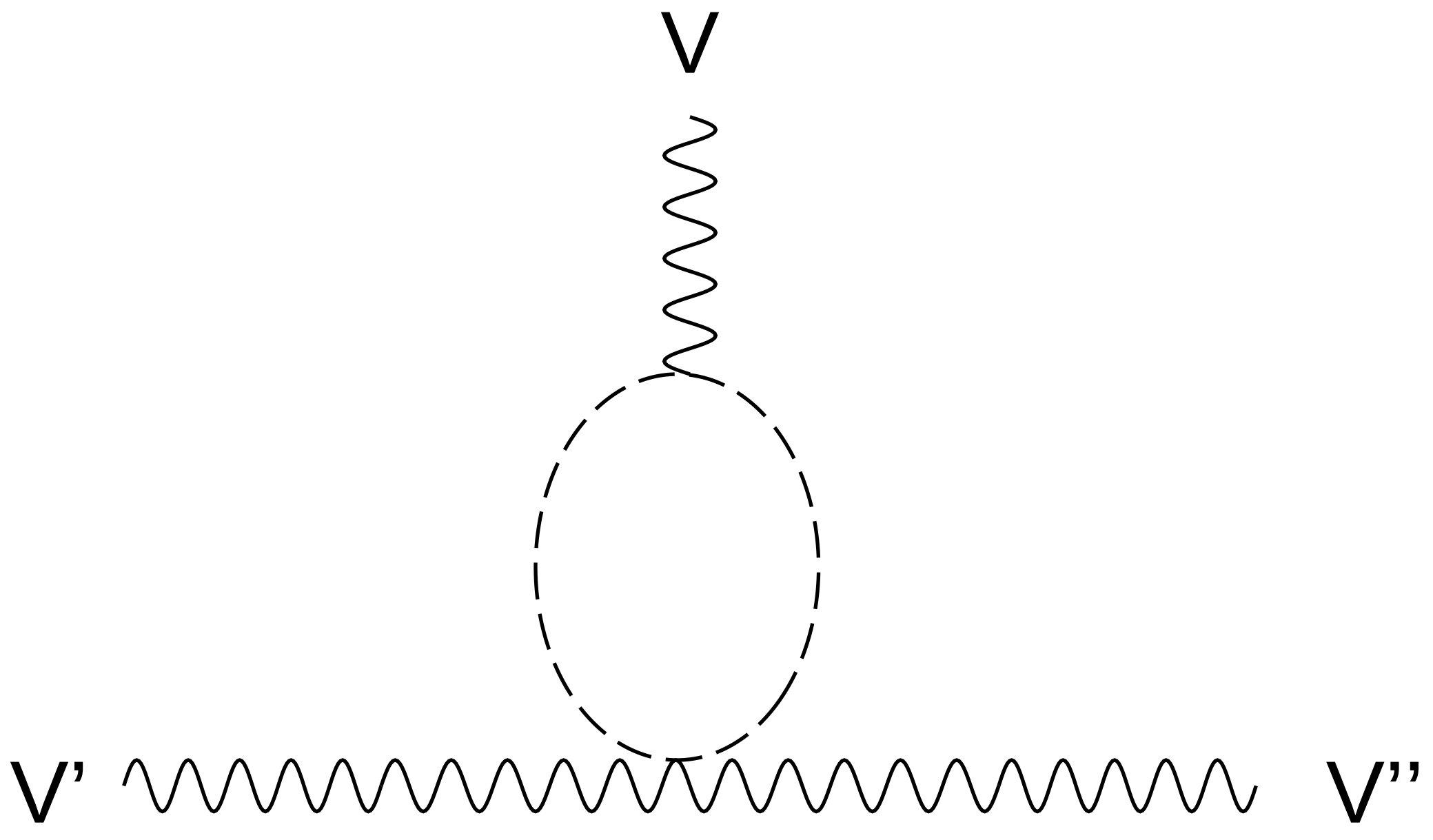

Figure 4

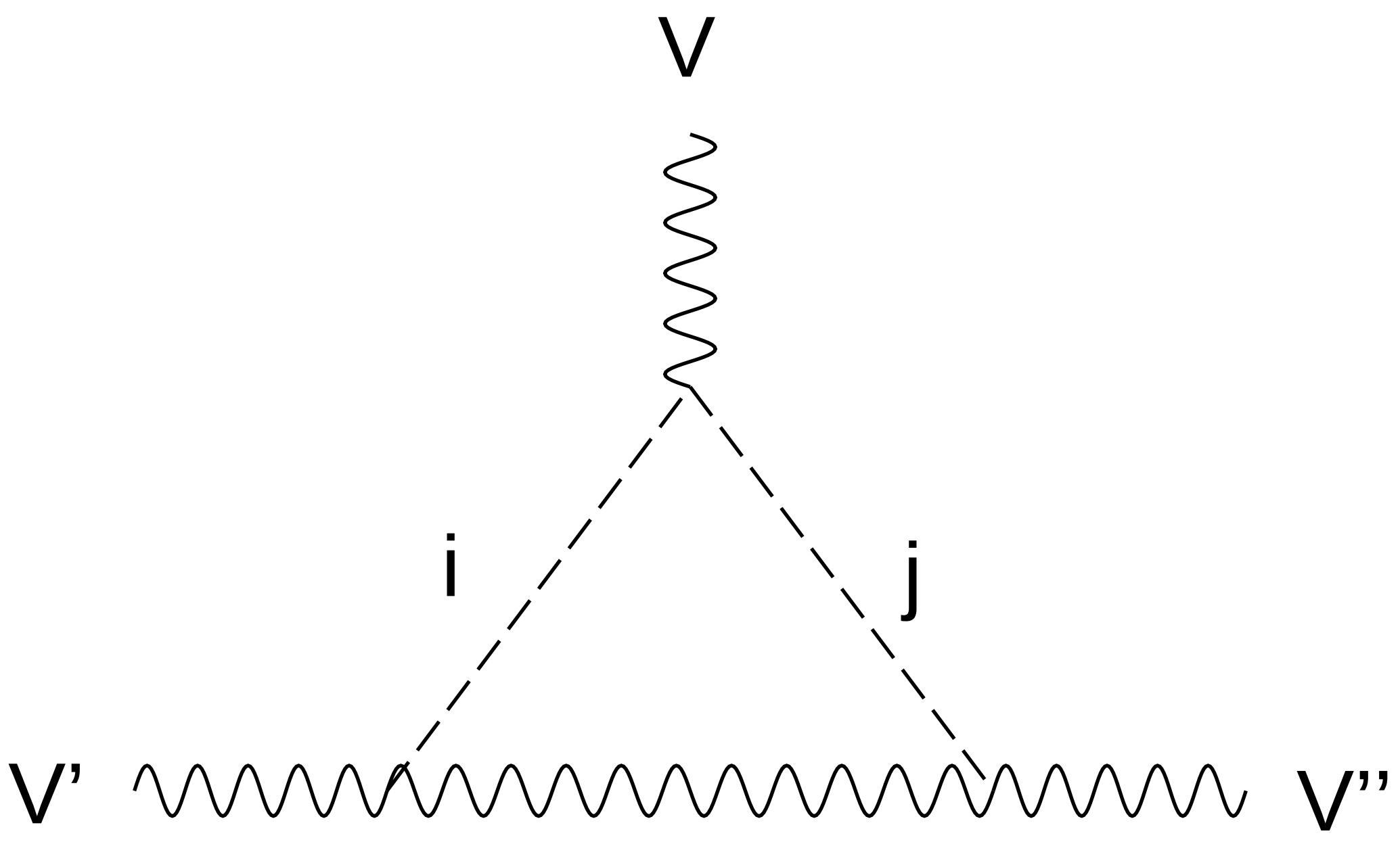

Figure 5

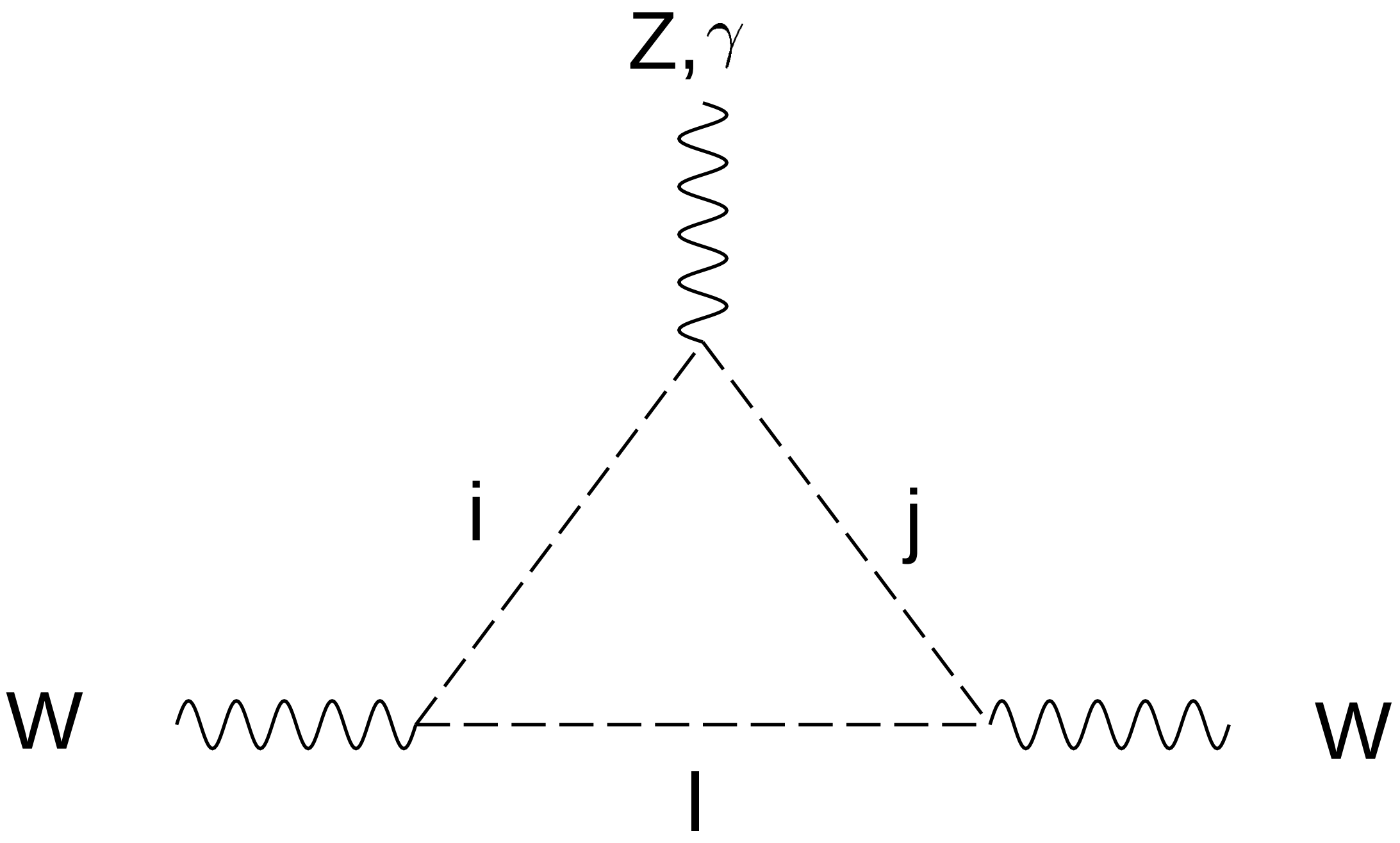

Figure 6

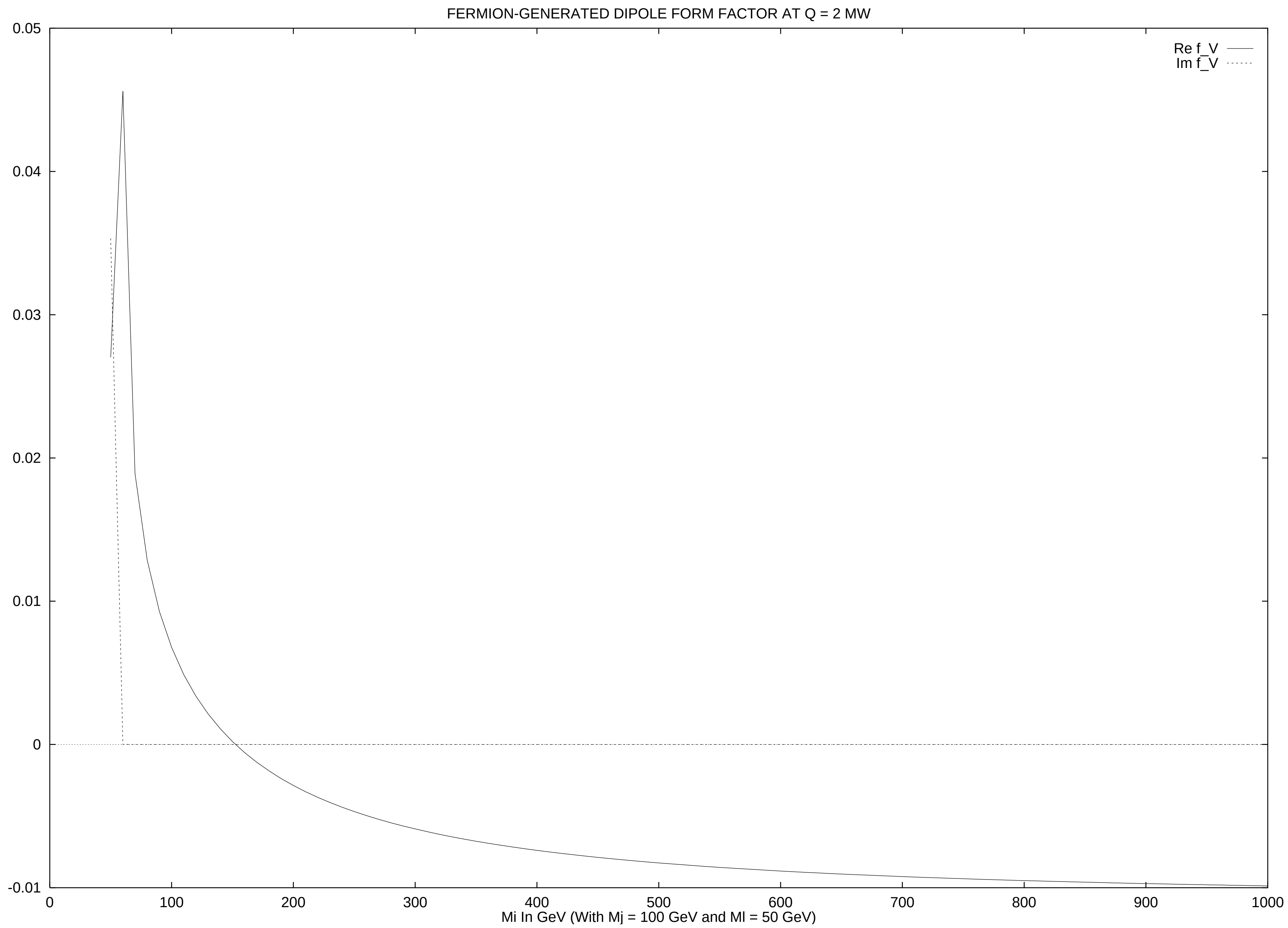

FERMION-GENERATED DIPOLE FORM FACTOR AT Q = 2 MW
Re f_V
Im f_V
0.05
0.04
0.03
0.02
0.01
0
-0.01
0
100
200
300
400
500
600
700
800
900
1000
Mi In GeV (With Mj = 100 GeV and Ml = 50 GeV)

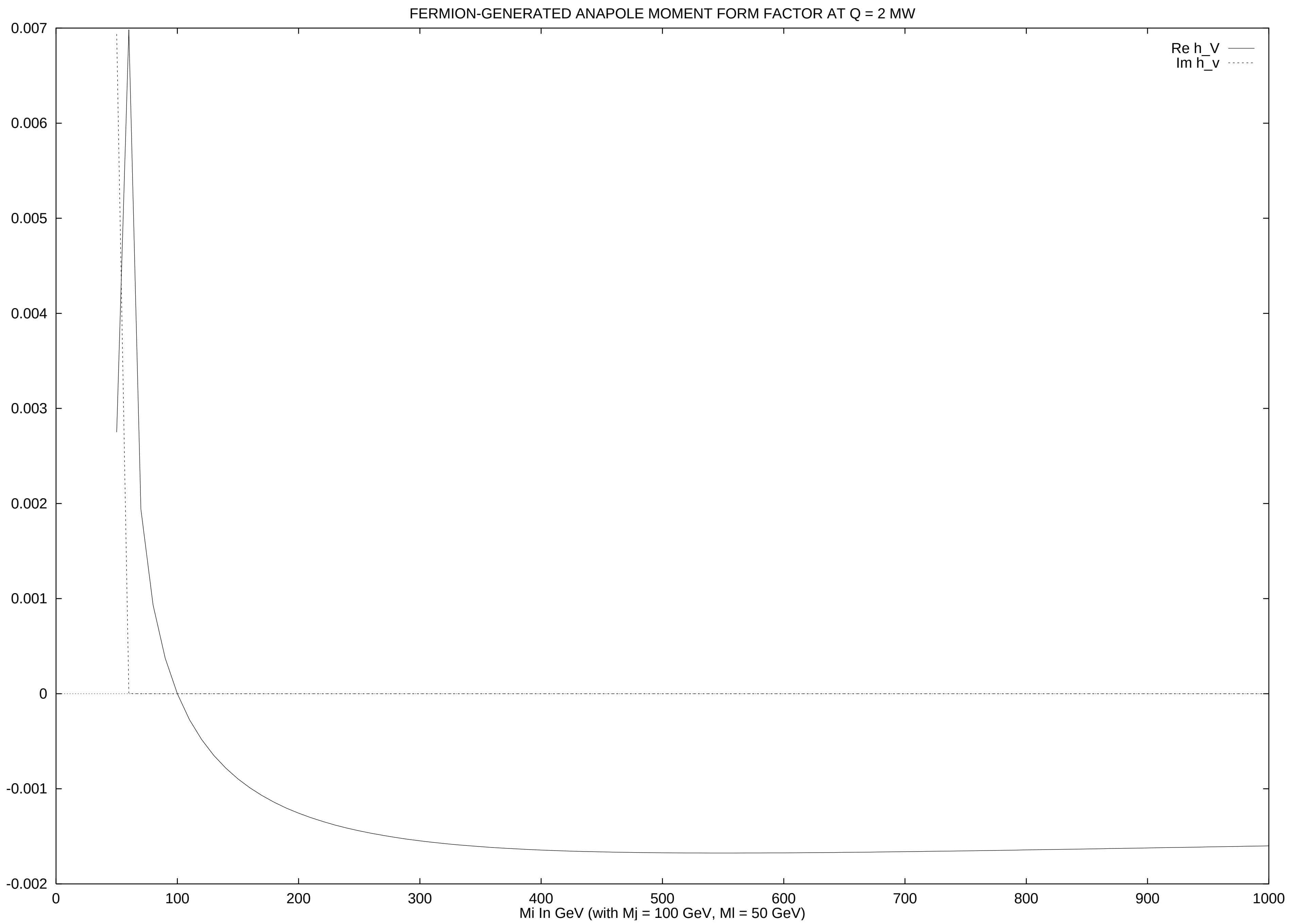
FERMION-GENERATED ANAPOLE MOMENT FORM FACTOR AT Q = 2 MW
Re h_V
Im h_v
0.007
0.006
0.005
0.004
0.003
0.002
0.001
0
-0.001
-0.002
0
100
200
300
400
500
600
700
800
900
1000
Mi In GeV (with Mj = 100 GeV, Ml = 50 GeV)

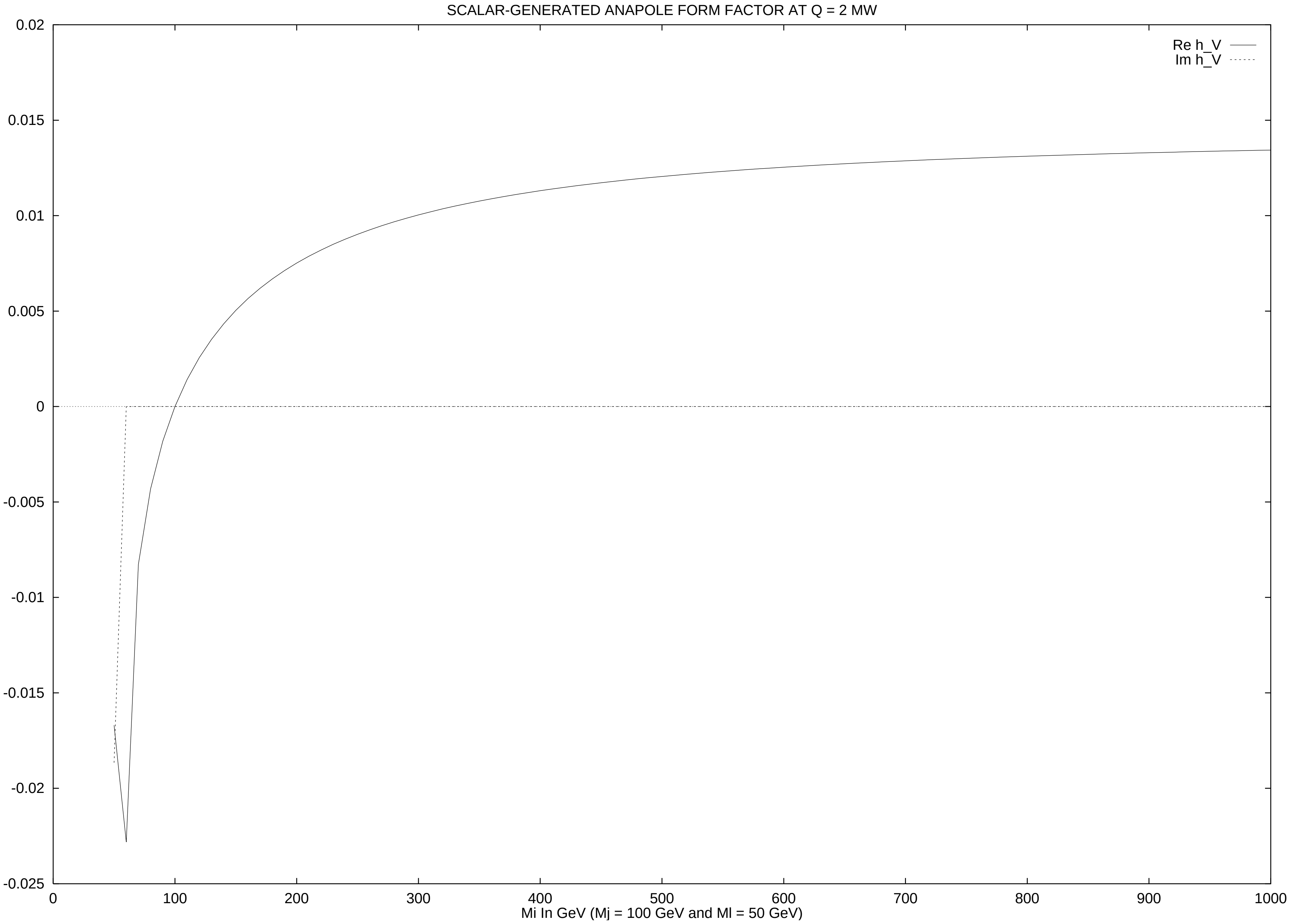
SCALAR-GENERATED ANAPOLE FORM FACTOR AT Q = 2 MW
Re h_V
Im h_V
0.02
0.015
0.01
0.005
0
-0.005
-0.01
-0.015
-0.02
-0.025
0
100
200
300
400
500
600
700
800
900
1000
Mi In GeV (Mj = 100 GeV and Ml = 50 GeV)

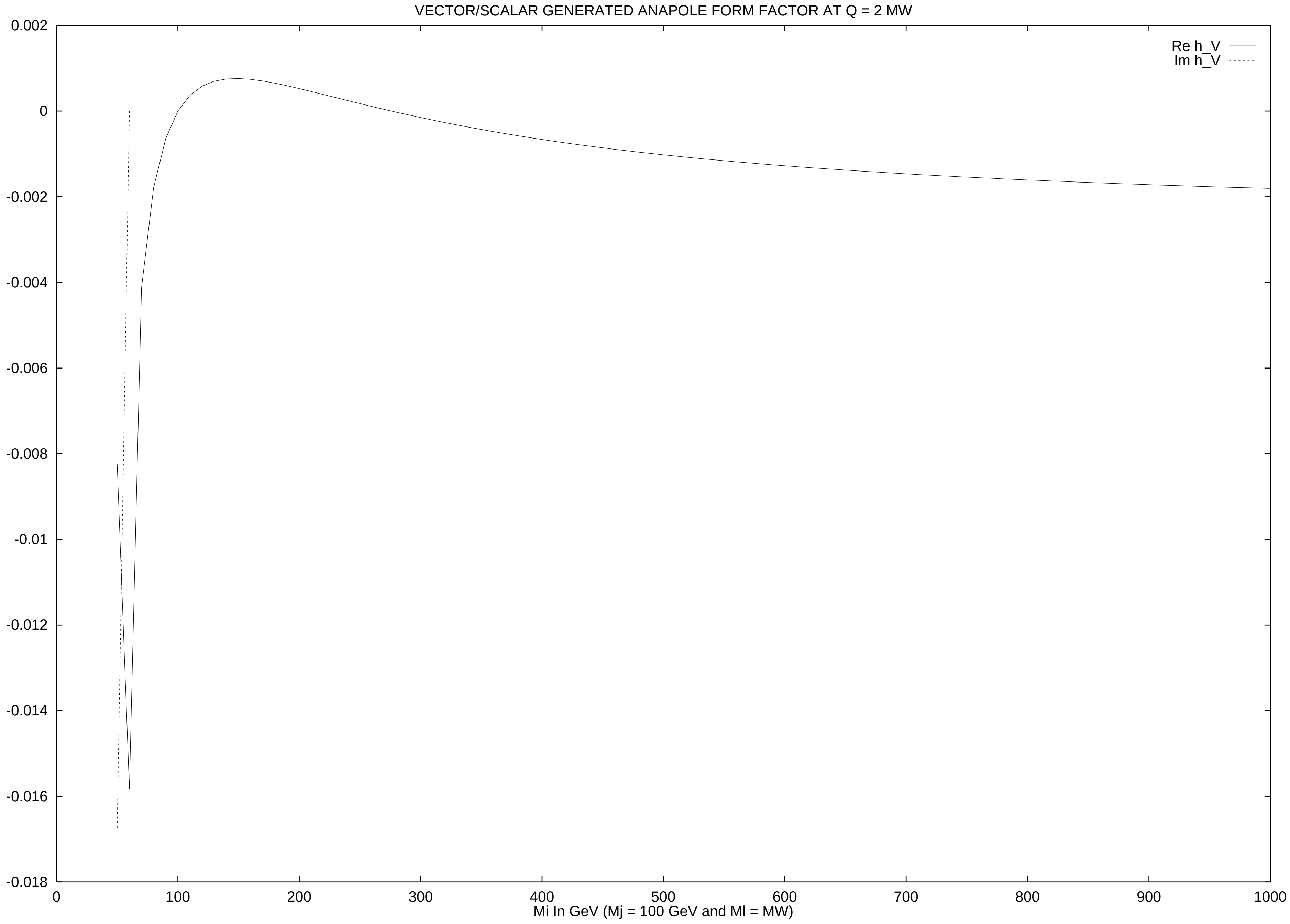
VECTOR/SCALAR GENERATED ANAPOLE FORM FACTOR AT Q = 2 MW
Re h_V
Im h_V
0.002
0
-0.002
-0.004
-0.006
-0.008
-0.01
-0.012
-0.014
-0.016
-0.018
0
100
200
300
400
500
600
700
800
900
1000
Mi In GeV (Mj = 100 GeV and Ml = MW)

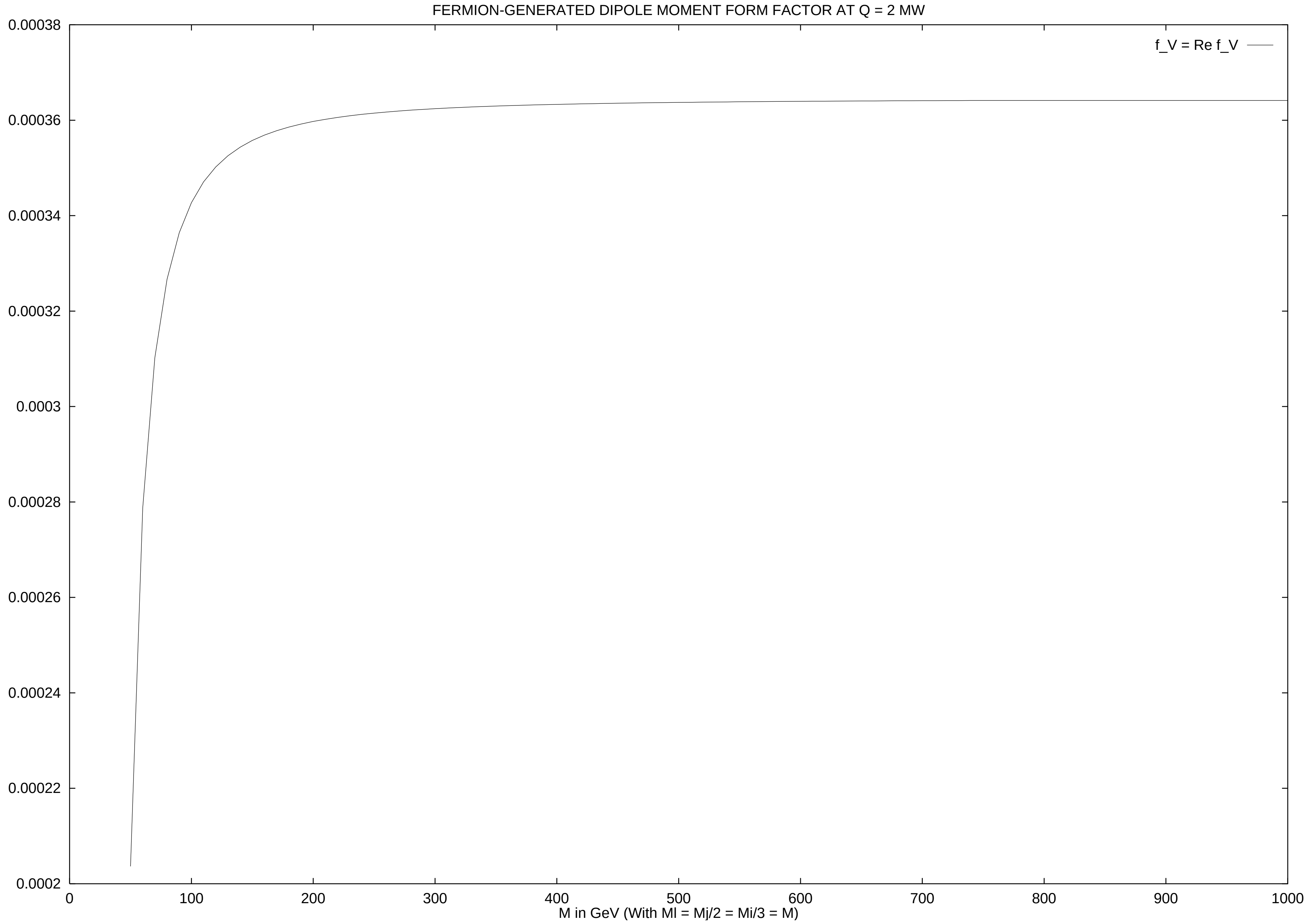

FERMION-GENERATED DIPOLE MOMENT FORM FACTOR AT Q = 2 MW
f_V = Re f_V
0.00038
0.00036
0.00034
0.00032
0.0003
0.00028
0.00026
0.00024
0.00022
0.0002
0
100
200
300
400
500
600
700
800
900
1000
M in GeV (With Ml = Mj/2 = Mi/3 = M)